\catcode`\@11
\documentclass[11pt]{article}
\usepackage{times,pifont}
\usepackage{amsmath,amssymb}
\usepackage{amsfonts,dsfont} 
\usepackage[mathscr]{eucal}
\usepackage[all]{xy}
\usepackage{cite}
\newcommand\preprint[1]{\vspace{-1in}\vtop{\null\hfill
\parbox[t]{1.6in}{\small\sc #1\\\null}}
\vskip .5in\bigskip\normalfont}
\newcommand{\nc}{\newcommand}
\newcommand{\fn}{\footnote}
\numberwithin{equation}{section}
\newcommand{\bec}{\begin{center}}
\newcommand{\eec}{\end{center}}
\newcommand{\bra}[1]{\left\langle\left. #1\right.\right\vert}
\newcommand{\ket}[1]{\left\vert\left. #1\right.\right\rangle}
\def\appendix{{\section{Appendix}}\let\appendix\section%
        {\setcounter{section}{0}
        \gdef\thesection{\Alph{section}}}\section}
\nc{\isk}[1]{|\; #1\; \rangle\rangle}
\nc{\isb}[1]{\langle\langle\; #1\; |}
\nc{\Isk}[1]{\Big\rvert\; #1\; \Big\rangle\Big\rangle}
\nc{\Isb}[1]{\Big\langle\Big\langle\; #1\; \Big\lvert}
\newcommand{\be}{\begin{equation}}
\newcommand{\ee}{\end{equation}}
\newcommand{\refb}[1]{(\ref{#1})}
\nc{\bbm}{\begin{bmatrix}}
\nc{\ebm}{\end{bmatrix}}

\nc{\beb}{\bigl[\begin{smallmatrix}}
\nc{\eeb}{\end{smallmatrix}\bigr]}
\nc{\bep}{\begin{pmatrix}}
\nc{\eep}{\begin{pmatrix}}

\nc{\etal}{\eta_L}
\nc{\etar}{\eta_R}
\nc{\imply}{\Rightarrow}
\nc{\dn}{\downarrow}
\nc{\chat}{\hat{c}}
\nc{\N}{{\sf N}}
\nc{\NN}{{\cal N}}
\nc{\XX}{{\cal X}}
\nc{\SSS}{{\cal S}}
\nc{\NS}{\NN\SSS}
\nc{\RA}{{\cal R}}
\nc{\OO}{{\cal O}}
\nc{\PP}{{\cal P}}
\nc{\SP}{{\sf P}}
\nc{\UU}{{\cal U}}
\nc{\CY}{\text{CY}}
\nc{\cp}{{\cal P}^2}
\nc{\BP}{\mathds{P}}
\nc{\map}{\longmapsto}
\nc{\ox}{\otimes}
\nc{\x}{\times}
\nc{\w}{\wedge}
\nc{\W}{\bigwedge}
\nc{\p}{\partial}
\nc{\pbar}{\bar{\partial}}
\nc{\MM}{{\sf M}}
\nc\wt{\widetilde}
\nc{\wh}{\widehat}
\nc{\vp}{\varphi}
\nc{\ep}{\epsilon}
\nc{\vep}{\varepsilon}
\nc{\RR}{\mathbf{R}}
\nc{\RE}{{\cal R}}
\nc{\Z}{\mathds{Z}}
\nc{\ZZ}{\mathds{Z_2}}
\nc{\RP}{\mathbb{R}\mathbb{P}}
\nc{\viz}{{\em{viz. }}}
\nc{\eg}{{\em{e.g. }}}
\nc{\ie}{{\em{i.e. }}}
\renewcommand\bar[1]{\smash{\overline{#1}}}
\nc{\ub}{\underbrace}
\nc{\ob}{\overbrace}
\nc{\C}{\mathds{C}} 
\nc{\zbar}{\bar{z}}
\nc{\wbar}{\bar{w}}
\nc{\noi}{\noindent}
\nc{\eights}{\boldsymbol{8}_{\boldsymbol{\rm s}}}     
\nc{\eightc}{\boldsymbol{8}_{\boldsymbol{\rm c}}}     
\nc{\eightv}{\boldsymbol{8}_{\boldsymbol{\rm v}}}     
\nc{\sla}{(S_L)_{\alpha}}
\nc{\slb}{(\bar{S}_L)^{\dot{\alpha}}}
\nc{\sra}{(S_R)_{\beta}}
\nc{\srb}{(\bar{S}_R)^{\dot{\beta}}}
\nc{\wil}{(-1)^{F_L}}           
\nc{\wi}{(-1)^{F}}              
\nc{\wir}{(-1)^{\wt{F}}}         
\nc{\wip}{(-1)^{F'}}
\nc{\wirp}{(-1)^{\wt{F'}}}
\nc{\II}{{\cal I}}
\nc{\g}{\mathfrak{g}}
\nc{\JJ}{{\cal J}}
\nc{\TO}{{\cal T}}
\nc{\KB}{{\cal K}}
\nc{\AN}{{\cal A}}
\nc{\MS}{{\cal M}}
\nc{\CC}{{\cal C}}
\nc{\HH}{{\cal H}}
\nc{\GG}{{\cal G}}
\nc{\BB}{{\cal B}}
\nc{\one}{{\mathbf{1}}}
\nc{\tq}{\widetilde{q}}
\nc{\hq}{\wh{q}}
\nc{\Lbar}{\bar{L}}
\nc{\dps}{\displaystyle}
\nc{\Jbar}{\bar{J}}
\nc{\ibar}{\bar{i}}
\nc{\jbar}{\bar{j}}
\nc{\YY}{{\cal Y}}
\nc{\jhat}{\wh{j}}
\nc{\pjhat}{\wh{j^{\prime}}}
\nc{\nhat}{\wh{n}}
\nc{\pnhat}{\wh{n^{\prime}}}
\nc{\shat}{\wh{s}}
\nc{\jtilde}{\wt{j}}
\nc{\pjtilde}{\wt{j^{\prime}}} 
\nc{\ntilde}{\wt{n}}
\nc{\pntilde}{\wt{n^{\prime}}}
\nc{\stilde}{\wt{s}}
\nc{\pstilde}{\wt{s^{\prime}}}
\nc{\jprime}{j^{\prime}}
\nc{\nprime}{n^{\prime}}
\nc{\sprime}{s^{\prime}}
\nc{\Tr}{{\rm Tr}}
\nc{\ul}{\underline}
\nc{\sigmahat}{\mathcal{R}}
\nc{\half}{\frac{1}{2}}
\nc{\inthree}{\frac{1}{3}}
\nc{\infour}{\frac{1}{4}}
\nc{\insix}{\frac{1}{6}}
\nc{\inseven}{\frac{1}{7}}
\nc{\ineight}{\frac{1}{8}}
\nc{\intwelve}{\frac{1}{12}}
\nc{\sqtwo}{{1\over\sqrt 2}}
\nc{\EV}{U(t)}
\nc{\ns}{{\rm NS}}
\nc{\NSNSU}{{\rm NSNS,U}}
\nc{\NSNST}{{\rm NSNS,T}}
\nc{\RRU}{{\rm RR,U}}
\nc{\RRT}{{\rm RR,T}}
\nc{\RT}{{\rm R,T_k}}
\nc{\NSNS}{{\rm NSNS}}
\nc{\RaRa}{{\rm RR}}
\nc{\U}{{\rm U}}
\nc{\T}{{\rm T}}
\nc{\xcap}{{\rm Xcap}}
\nc{\GSO}{{\rm GSO}}
\nc{\bp}{\bar{p}}
\nc{\bi}{\bar{i}}
\nc{\bz}{\bar{z}}
\nc{\IM}{{\rm Im}}
\nc{\re}{{\rm Re}}
\nc{\zero}{\Big|_{{\rm zero}}}
\nc{\nonzero}{\Big|_{{\rm Nonzero}}}
\nc{\sqthree}{{1\over\sqrt 3}}
\nc{\onethird}{{1\over 3}}
\nc{\wl}{\wh{l}}
\nc{\wpp}{\wh{p}}
\nc{\wrr}{\wh{r}}
\nc{\ws}{\wh{s}}

\newcommand\norder{\genfrac{}{}{0pt}{3}{\circ}{\circ}}
\def\rt{\longrightarrow}
\begin{document}
\title{\preprint{hep-th/0602135\\HIP-2006-08/TH\\IP/BBSR/2006-7}  
Fractional Branes in Non-compact Type IIA Orientifolds }
\author{ 
Jaydeep Majumder,
\thanks{jaydeep dot majumder at helsinki dot fi} \\ 
\small  Helsinki Institute of Physics, 
University of Helsinki 
 P O Box 64, FIN 00014, Finland.\\ 
{}\\
Subir Mukhopadhyay
\thanks{subir at iopb dot res dot in}\\
\small Institute of Physics, \
Bhubaneswar 751~005, India.
\\ \&
\\
Koushik Ray \thanks{koushik at iacs dot res dot in} \\
\small Department of Theoretical Physics,\
Indian Association for the Cultivation of Science\\
\small Calcutta 700~032. India.
}
\date{}
\maketitle
\thispagestyle{empty}
\begin{abstract}
We study fractional D-branes in the Type-IIA theory
on a non-compact orientifold of the orbifold $\C^3/\Z_3$
in the boundary state formalism.
We find that the fractional D0-branes of the
orbifold theory become unstable
due to the presence of a tachyon, while there
is a stable D-instanton whose tachyon
gets projected out. We propose that
the D-instanton is obtained after
tachyon condensation.
We evidence this by calculating the Whitehead group of the Abelian category 
of objects corresponding to the boundary states
as being isomorphic to $\Z_2$. 
\end{abstract}
\clearpage
\section{Introduction and Summary}\label{s0}
D-branes have become a sine qua non for the studies of non-perturbative 
string theory. The description of D-branes in terms of open strings 
makes it possible to treat them within the scope of a boundary conformal field theory 
(BCFT). Since the BCFT does not rely upon spacetime supersymmetry, this formulation is 
well-suited for treating BPS as well as non-BPS states on the same footing. 
D-branes have been studied on a variety of singular and non-singular 
target spaces using BCFT \cite{hep-th/9612126,hep-th/9712230,
hep-th/9906242,hep-th/0009072,hep-th/0006224,
hep-th/0102146,hep-th/9906200,hep-th/9912147,hep-th/9912188,Govindarajan:1999js,
hep-th/0007075,hep-th/0010196,
hep-th/0010217,hep-th/0010223,hep-th/0401135,hep-th/0504164,hep-th/0002037,
hep-th/0003263,Billo:2000yb,Gaberdiel:2000fe,Craps:2001xw,Klemm:2005gc,Maiden:2006qe,cetic}.
D-branes in the Type-II theories on non-compact orbifolds constitute an interesting
class of such theories \cite{hep-th/9603167,hep-th/9610140,hep-th/9704151,
hep-th/9612126,hep-th/9712230,hep-th/9906242}.
The stable states in the spectrum of such theories have been identified with D-branes 
wrapped on various supersymmetric cycles of the target space. One of the purposes of 
the present article is to go beyond orbifold backgrounds to the more interesting
orientifold backgrounds \cite{hep-th/9804208,hep-th/0204089,BZ,GP,GJ} and examine 
the spectrum of stable D-branes.

While open strings arise as fluctuations of D-branes, the latter can be thought of as
a geometric description of the gauge degrees of freedom ensuing from the terminal
points of the former. Hence the interpretation of the states in the spectrum of open
strings in terms of D-branes becomes transparent when viewed in the closed string
channel. This is brought out through boundary states, which incorporate the boundary
conditions of open strings in the closed string language. A formulation of BCFT is in
terms of such boundary state. 
We consider D-branes in the Type-IIA string theory on an orientifold of the 
three-dimensional orbifold $\C^3/\Z_3$
using the boundary state formulation \cite{CALLAN,ISHIBASHI,DFMS}. 
The orientifold reduces spacetime supersymmetry further compared to 
the parent orbifolds 
\cite{hep-th/9906242,hep-th/0102146,hep-th/0009072,DelaOssa,
hep-th/9908130,hep-th/9912204,hep-th/0211059,hep-th/0401040,
hep-th/0404254,hep-th/9912218,hep-th/0305108,hep-th/0306257,hep-th/0401137,
hep-th/0511100,hep-th/0304209,hep-th/0305021}, 
rendering the standard 
machinery of the $(2,2)$ theories unavailable. Nevertheless, D-branes on 
certain non-compact three-dimensional orientifolds have been studied earlier,
most which, however, dealt with the Type-IIB string theory 
\cite{hep-th/9910109,hep-th/9812135}. 
In the present article we consider the Type-IIA theory on the orientifold, which  
is a cousin of a certain asymmetric orbifold with magnetic fluxes on D-branes, of the Type-IIB theory via T-duality 
\cite{hep-th/0003024}, which makes it rather different from the earlier
analyses as we shall observe\fn{For example, 
the Type IIA orientifold model we discuss, is not T-dual to the Type-IIB 
oreintifold on $\C^3/\Z_3$ of \cite{hep-th/9606169}.}.

Prior to orientifolding the stable D-brane 
configurations are the fractional D0-branes\cite{hep-th/9612126,hep-th/9712230}. 
In the present example
the D0-branes are inflicted with tachyonic instabilities after orientifolding. 
However, we find stable D-instanton configurations in the model which are otherwise 
absent in the parent orbifold theory. The unavailability of the $(2,2)$ machinery poses
a major hurdle in arriving at a geometric interpretation of these states. 
Nevertheless, due to the invariance of the boundary states of the D0-brane under the
orientifolding operation we can first consider the geometric objects in the parent
orbifold theory. Then, by lifting the orientifold action on these objects as an
automorphism squaring up to the identity leads to the
geometric entities present in the orientifold theory. We take this approach in this
article. 

In the parent orbifold theory the D-brane configurations 
are identified with objects in the derived category
of an Abelian category, where the latter
can be identified with the category of coherent 
sheaves on the blown-up orbifold in the large volume limit 
and with the category of irreducible representation of an associate quiver in the
so-called orbifold regime \cite{hep-th/0011017}. 
D-brane configurations in the Type-IIA and Type-IIB theories 
can be identified as the elements of the $K^1$ and $K^0$ groups of 
the Abelian category and its equivariant descendants on spacetime orbifolds. 
On orientifolds the stable D-branes are classified by higher K-theoretic charges.
In the present case, the D-brane configurations correspond to the 
elements of the Whitehead group, $K^1$, of the Abelian category 
associated with the orientifold space. Assuming that the orientifold operation induces
an automorphism on the objects of the Abelian category that squares up to the identity, 
that is, 
the boundary states, we calculate the Whitehead group of the Abelian category. 
We find that the Whitehead group is isomorphic to $\Z_2$, which we identify as the
charge of the D-instanton. 
 
The paper is organized as follows. 
In the following section we describe the  
orientifold on which we compactify  the Type-IIA theory.
Then in the two subsequent sections we construct 
the crosscap state and the D-brane boundary state 
respectively in the BCFT formulation. 
Next we calculate various one-loop amplitudes. 
We show that the D0-branes are plagued 
with the tachyonic instability and that the D-instanton
gives rise to a stable configuration, instead. 
The following section deals with the K-theoretic analysis. 
Finally we conclude with discussion of the results. Some of the 
useful formulas in our notation and conventions have been relegated to the Appendix.


\section{The orientifold}\label{s1}
Let us begin our discussion by describing the theory under consideration. 
In this section we first discuss the orientifold action and then the 
spectrum of the massless closed string states. This analysis provides 
the space-time fields present in the theory.
\subsection{Orientifold action}
\label{secoa}
We consider Type--IIA theory in the light-cone gauge on the orientifold
\cite{hep-th/9912204} $\C^3/\GG$ with $\GG = 
(\Omega\cdot\sigmahat\cdot\wil)\otimes G$, 
where $\GG$ is a discrete group with both geometric and 
non-geometric parts. The first piece of $\GG$ in parentheses 
refers to the diagonal group isomorphic to $\Z_2$ obtained 
by combining three $\Z_2$ groups. Of these, $\Omega$, isomorphic to $\Z_2$,
acts on the world-sheet fields by reversal of parity,
$\Omega:\sigma\map\pi - \sigma$, where $\sigma$ denotes the spatial 
coordinate of the world-sheet. 
The anti-holomorphic involution $\sigmahat$, also isomorphic to $\Z_2$,
acts by complex conjugation 
$\sigmahat : Z^i \map\bar{Z}^i$, on the complex
bosonic fields of the world-sheet theory, whose zero-modes are identified with the complex
coordinates of the $\C^3$. This is tantamount to a reflection
of three of the corresponding real coordinates.
These are further accompanied with
$\wil$, which changes the sign of 
the left-moving space-time fermions in order for making $\GG$ into
a symmetry group of the Type--IIA theory. Finally, $\GG$ contains
a cyclic group $G$ isomorphic to $\Z_N$ and generated by $g$ acting as,
\begin{equation} \label{e1}
g: Z^i\map e^{2\pi ikv_i} Z^i,
\end{equation} 
on the complex coordinates of $\C^3$
and similarly on their fermionic counterparts $\Psi^i$, $i=1,2,3$.
While much of the discussion in the sequel 
remain unaltered for any odd integer $N$, 
we shall restrict ourselves to the specific case of $N=3$ for  
simplicity. Thus, we have $k = 0,1,2$  and 
$\vec{v} = (1/3, 1/3, -2/3)$.
The compact cousin of this model
was discussed in \cite{hep-th/9912204}.
The combined action of $\Omega$
and $\wil$ on the worldsheet fermions is given by,
\begin{equation} \label{e3}
\Omega\cdot\sigmahat :\Psi^i(\sigma)\map\bar{\wt{\Psi^i}}(\pi-\sigma),
\end{equation} 
where a tilde designates a  right-moving field. Let us also note that 
by abuse of notation we denote the groups $\Omega$, $\sigmahat$ and $\wil$
as well as their respective generators by the same symbols. 

We have described the action of $\GG$ on the fields of the theory, which can 
be used to describe the action on the corresponding oscillators in their
mode expansions, as described in Appendix.
Let us now discuss its lift to the states of the theory.
The unique ground state $\ket{0}_{\text{NS}}$ in the NS-sector remains unaffected.
Let us write the left- and the right-moving Rammond ground states,
hereafter referred to as the R-ground states,
as $\ket{\mathbf{s}}_L=\ket{s_0,\vec{s}}_L=\ket{s_0,s_1,s_2,s_3}_L$
and $\ket{\mathbf{\wt{s}}}_R=\ket{\wt{s}_0,\vec{\wt{s}}}_R=
\ket{\wt{s_0},\wt{s_1},\wt{s_2},\wt{s_3}}_R$
respectively, where $s_a$, $\wt{s_a}=
\pm\frac{1}{2}$ for $a=0,1,2,3$.
In our convention, the left-moving R-ground states, transforming as $\eights$ 
under the Poincar\'e group $SO(8)$
of the space-time transverse to the 
light-cone are chosen to be the ones with 
a  even number of $-{1}/{2}$'s
while the right-moving R-ground states, transforming as $\eightc$ 
under the $SO(8)$, are taken to be the ones in which an odd number 
of $-{1}/{2}$'s occur. 
The action of $\GG$ on the various R-ground states is given by,
\begin{gather}
g^k: \ket{s_0,\vec{s}}_L\map
e^{2\pi ik{\vec v}\cdot{\vec s}}\,
\ket{s_0,\vec{s}}_L, \quad
\ket{\wt{s_0},\vec{\wt{s}}}_R\map
e^{2\pi ik{\vec v}\cdot{\vec{\wt{s}}}}\,
\ket{\wt{s_0},\vec{\wt{s}}}_R; \notag\\ 
\label{e8}
\wil :\ket{s_0,\vec{s}}_L\map -\ket{s_0,\vec{s}}_L, \quad 
\ket{\wt{s_0},\vec{\wt{s}}}_R\map
\ket{\wt{s_0},\vec{\wt{s}}}_R; \\
\Omega\cdot\sigmahat : 
\ket{s_0,\vec{s}}_L\map
\ket{{s_0},-\vec{{s}}}_R, \quad 
\ket{\wt{s_0},\vec{\wt{s}}}_R\map
e^{\pi i(\wt{s_0}+\wt{s_1}+\wt{s_2}+\wt{s_3})}
\ket{\wt{s_0},-\vec{\wt{s}}}_L = - \ket{\wt{s_0},-\vec{\wt{s}}}_L,\notag
\end{gather}
as the sum 
$\dps\sum_{a=0}^3 \wt{s_a}$ is odd for states belonging to $\eightc$, resulting into 
\begin{equation} \label{e11}
\g(k)\equiv
\Omega\cdot\sigmahat\cdot\wil\cdot g^k:
\ket{s_0,\vec{s}}_L\ox 
\ket{\wt{s_0},\vec{\wt{s}}}_R\map 
- e^{2\pi ik\vec{v}\cdot(\vec{s}+\vec{\wt{s}})} 
\ket{\wt{s_0},-\vec{\wt{s}}}_L\ox
\ket{s_0,-\vec{s}}_R,
\end{equation} 
after taking into account the minus sign that arises in 
exchanging fermions. In the above formulas an $s$ in the right-moving 
state or an $\wt{s}$ in the left-moving one is interpreted as their 
respective numerical values\footnote{We can also label the massless states in 
NS-sectors by their $SO(8)$ weights. Thus, if $s_a = \wt{s_a} = (\underbrace{\pm 1,0,0,0})$, 
where $\underbrace{}$ denotes all possible permutations, such states belong to 
$\eightv$ of $SO(8)$. Since $\sum_a s_a  =\sum_a \wt{s_a} =\text{odd}$ holds for a state 
belonging to $\eightv$, the last equation of \refb{e8} holds. Thus the action of the 
orientifold group on massless NS-NS massless states is given by eqn
\refb{e11} but {\em without} the $-$ sign, since $\wil$ has no action this time.}.  

Finally, the left moving and right moving world sheet 
fermion number operators are defined as
\begin{equation} 
\label{e4a}
\begin{split}
\wi\ket{s_0,s_1,s_2,s_3}_L &= -e^{\pi i(s_0+s_1+s_2+s_3)}\ 
\ket{s_0,s_1,s_2,s_3}_L
\\
\wir\ket{\wt{s_0},\wt{s_1},\wt{s_2},\wt{s_3}}_R &= 
-e^{\pi i(\wt{s_0}+\wt{s_1}
+ \wt{s_2}+\wt{s_3})}\ 
\ket{\wt{s_0},\wt{s_1},\wt{s_2},\wt{s_3}}_R\,,
\end{split}
\end{equation} 
So, we can choose the 
GSO-projection operators as in the untwisted sector as,
\begin{equation} \label{GSO}
\begin{split}
\mathcal{P}_U =
\begin{cases}
 \Big(\frac{1 + \wi}{2}\Big)\oplus
\Big(\frac{1 + \wir}{2}\Big),\quad\text{in the NS-NS sector}\\
\Big(\frac{1 - \wi}{2}\Big)\oplus
\Big(\frac{1 + \wir}{2}\Big),\quad\text{in the R-R sector}.
\end{cases}
\end{split}
\end{equation} 
In the above expression and in what follows a subscript $U$ is taken to
designate a quantity in the untwisted sector, while one in the twisted 
sector will be designated by $T$.
\subsection{Closed string spectrum}\label{s2}
Let us now describe the spectrum of massless closed string states 
in four dimensions that survive the orientifolding described above. 
\subsection*{Untwisted sector}
The untwisted sector corresponds to $k=0$ in equation \refb{e8}. 
In NS-NS sector, first, we have the four dimensional 
graviton, $g_{\mu\nu}$ and dilaton, $\phi$, whereas 
the four dimensional part of the $B$-field 
is projected out. However, few components of 
the metric and $B$-field along the ``internal'' 
$\C^3$ directions survive.
In terms of the oscillators, these states 
are given by  six $\GG$-invariant combinations
of states, namely,
\begin{equation} 
\label{e5}
(\Psi^i_{-\half}\bar{\wt{\Psi^j}}_{-\half} + 
\Psi^j_{-\half}\bar{\wt{\Psi^i}}_{-\half})
\ket{0}_{\NSNS}, \quad i \geq j
\end{equation} 
where $i,j=1,2,3$ and $\ket{0}_{\NSNS}$ denotes the ground state in the NS-NS sector,
giving rise to three chiral multiplets in four dimensions\fn{We could have written down these states in terms of the $SO(8)$ weight notation, like we did for R-R case.}.

Let us now consider the untwisted states in the R-R sector. 
From equation \refb{e11} we see that 
under $\Omega\cdot\sigmahat$ the state $\ket{s_0,\vec{s}}$ goes to 
$\ket{\wt{s}_0,-\vec{\wt{s}}}$ in the left-moving sector and similarly for the 
right-moving ones. Thus, the states with 
$\vec{s}+\vec{\wt{s}}=0$ flip sign under $\GG$ and hence 
go away from the spectrum. However, certain linear combinations of 
those with $\vec{s}+\vec{\wt{s}}\neq 0$,
but satisfying ${\vec v}.(s+\wt{s})=0, 1$  survive and can be rearranged 
in the following seven independent states,
\begin{equation} \label{ecurr}
\begin{split}
\ket{++--}_L \ox \ket{+++-}_R 
&- 
\ket{+--+}_L \ox \ket{+-++}_R, \\
\ket{++--}_L \ox \ket{++-+}_R 
&- 
\ket{+-+-}_L \ox \ket{+-++}_R, \\
\ket{+-+-}_L \ox \ket{+++-}_R 
&- 
\ket{+--+}_L \ox \ket{++-+}_R, \\
\ket{-+-+}_L \ox \ket{-+--}_R 
&- 
\ket{--++}_L \ox \ket{--+-}_R,\\
\ket{-+-+}_L \ox \ket{---+}_R 
&- 
\ket{-++-}_L \ox \ket{--+-}_R, \\
\ket{--++}_L \ox \ket{---+}_R 
&- 
\ket{-++-}_L \ox \ket{-+--}_R, \\
\ket{++++}_L \ox \ket{-+++}_R 
&-
\ket{----}_L \ox \ket{+---}_R.
\end{split}
\end{equation} 
First six of these constitute three chiral multiplets,
while the last one joins the four-dimensional dilaton to form one more.
Thus, we have four chiral multiplets from the R-R sector in total.
\subsection*{\bf Twisted sector}
The massless closed string states in the twisted sector 
corresponding to $k=1,2$ in equation \refb{e8} are obtained
similarly. The GSO-projector in the twisted
sector, $\mathcal{P}_T$ is identical to the
GSO-projector in untwisted sector, 
$\mathcal{P}_T = \mathcal{P}_U$. 
Among  the GSO-invariant states a  single one
from each of the twisted NS-NS and R-R 
sectors survive the orientifolding. These are
\begin{equation} \label{e19}
\left[\ket{0,{1\over 3},{1\over 3},{1\over 3}}_L
\ox\ket{0,{1\over 3},{1\over 3},{1\over 3}}_R
+ 
\ket{0,-\frac{1}{3},-{1\over 3},-{1\over 3}}_L
\ox
\ket{0-{1\over 3}-{1\over 3}-{1\over 3}}_R\right], 
\end{equation} 
respectively, from the $k=1,2$ twisted NS-NS sector and 
\begin{equation} \label{e192}
\left[\ket{\half,-{1\over 6},-{1\over 6},-{1\over 6}}_L
\ox\ket{-\half,-{1\over 6},-{1\over 6},-{1\over 6}}_R 
+
\ket{-\half,{1\over 6},{1\over 6},{1\over 6}}_L\ox
\ket{\half,{1\over 6},{1\over 6},{1\over 6}}_R\right]
\end{equation} 
respectively, from the $k=1,2$ twisted R-R sector.
We thus get six chiral multiplets 
from the untwisted NS-NS sector and three from the untwisted R-R sector, adding up to
nine chiral multiplets in total\cite{hep-th/9912204}. 
The remaining state from equation \refb{ecurr} pairs up
with the the four-dimensional dilaton
to form an additional {\em dilaton multiplet}. 
Furthermore, the NS-NS and R-R twisted sectors together contribute one
chiral multiplet. As there is no vector multiplet,
all the D0-branes (including the fractional 
branes) are projected out by the orientifolding.
We shall confirm this from the analysis of the open string 
states in the following sections.
\section{The Crosscap state}\label{s3}
In order to study the D-branes in the presence of the orientifold plane
$O6$ in the model at hand
we need to study the open descendants of our model. This involves 
constructing the crosscap state, corresponding to the orientifold, and the 
boundary states corresponding to the D-branes. These are the states of 
open strings in the closed channel, also known as the direct or tree channel.

In this section we construct the crosscap state for the model at hand. 
The boundary states will be discussed in the following section.
A simplification in the construction of the crosscap state
in this model
ensues from the fact that 
the amplitudes can be obtained from crosscap states which does not
have a twisted sector\cite{hep-th/9908130}. In order to illustrate this
let us consider the Klein bottle amplitude $\mathcal{K}$. 
For a general $G\cong\Z_N$
the Klein bottle amplitude is given by 
\begin{equation}\label{kb}
\mathcal{K}=\frac{1}{N}\;\Tr
\Big[\Omega\cdot\sigmahat\cdot\wil\cdot(1+g+\cdots +g^{N-1})\mathcal{P}
q^{H_{c}}\Big]
\end{equation}
in the NS-NS or R-R sectors. 
However, from the action of the group $\mathcal{G}$ given in the 
previous section we find that the generators satisfy 
\begin{equation}\label{thetg}
\Theta g^k = g^{N-k}\Theta,
\end{equation}
where we introduced $\Theta = \Omega\cdot\sigmahat\cdot\wil$ for 
typographic ease.
Now, since the Hamiltonian is invariant under the action of the group
$G$, the energy eigenstates of the system are also eigenstates of the elements of  
$G$. Let us  fix a mutually orthogonal set of bases of such states, 
$\{\ket{m}|m=0,\ldots, N-1\}$, satisfying
\begin{equation} 
\label{eigeng}
g^k\ket{m} = e^{\frac{2\pi imk}{N}}\ket{m},
\end{equation} 
for $k=0,\ldots , N-1$. Hence the trace in equation \refb{kb} becomes 
a sum over the expectation values in these states. With these states we derive
\begin{equation}\label{ampone} 
\bra{m}\Theta g^k \mathcal{P}q^{H_{c}}\ket{m} 
= e^{\frac{2\pi i km}{N}}\bra{m}\Theta\mathcal{P}
q^{H_{c}}\ket{m},
\end{equation} 
and 
\begin{equation}\label{amptwo} 
\bra{m}g^{N-k}\Theta\mathcal{P}q^{H_{c}}\ket{m} 
= e^{-\frac{2\pi i km}{N}}\bra{m}\Theta\mathcal{P}
q^{H_{c}}\ket{m},
\end{equation} 
which have to be equal, by \refb{thetg}, thereby implying 
\begin{equation}
\label{sineq}
\bra{m}\Theta\mathcal{P} q^{H_{c}}\ket{m}
\sin\left(\frac{2\pi km}{N}\right)=0, 
\end{equation}
for non-zero $k$ and $m$. 
For $k=0$ or $m=0$ this equation collapses to an identity.
Consequently, the Klein bottle
amplitude \refb{kb} becomes
\begin{equation}
\begin{split}
{\cal K} &= \frac{1}{N}\sum\limits_{k,m=0}^{N-1} 
\langle m|\Theta g^k \mathcal{P} q^{H_c}|m\rangle \\
&= \frac{1}{N}\sum\limits_{k,m=0}^{N-1}e^{2\pi ikm/N} \langle m |\Theta\mathcal{P} q^{H_c}
|m\rangle,\\
\end{split}
\end{equation}
where we used the expression \refb{ampone}. Now, in \refb{sineq}, the first factor
being independent on $k$, for each non-zero $m$ we can find at least one value of  $k$
in the range $0 < k \leq N-1$, such that the sine factor is non-vanishing. Hence, the
amplitude $\bra{m}\Theta\mathcal{P} q^{H_{c}}\ket{m}$ vanishes for all non-zero $m$. 
\footnote{We thank the referee for raising this point.} 
Thus, finally, the Klein bottle amplitude becomes 
\begin{equation}
\begin{split}
 \mathcal{K} &= \frac{1}{N}\sum\limits_{k=0}^{N-1} \langle 0|\Theta \mathcal{P}
q^{H_c}|0\rangle \\
& = \langle 0|\Theta \mathcal{P} q^{H_c}|0\rangle .\\
\end{split}
\end{equation}
We have thus re-written the Klein bottle amplitude without the $g^k$-twisted open
string states, with no $g^k$ left in the expression. 
This implies that in the tree channel 
the Klein bottle amplitude can be obtained 
from the untwisted
crosscap state alone \cite{hep-th/9908130,hep-th/9912204, 
hep-th/9912204}. 
However, as we shall
discuss in the following section, the D-brane 
boundary states in the closed string picture do contain twisted pieces.

Let us now write down the equations satisfied by the 
crosscap state \cite{CALLAN, POLCHINSKI-CAI,ISHIBASHI}. 
These are obtained by twisting the periodic
boundary conditions of the closed string fields 
by the orbifold as well as orientifold action as,
\begin{gather}
(X^M(\sigma) - \g(k) X^M(\sigma+\pi)))
\ket{C_6;\eta} = 0,  \notag\\
\label{e29}
(\partial_{\tau}X^M(\sigma) + \g(k) \partial_{\tau} 
X^M(\sigma+\pi))\ket{C_6;\eta} = 0,  \\
(\psi^M(\sigma) + i\eta\,\g(k)\wt{\psi}^M(\sigma+\pi,0))
\ket{C_6;\eta} = 0,\notag
\end{gather}
valid at $\tau = 0$, 
where $M=0,\ldots , 9$, $\g(k)$ is as defined in equation \refb{e11}
and $C_6$ designates the $O6$-plane. 
Such a crosscap state was first discussed in 
\cite{hep-th/0105208,hep-th/0201037}
\fn{For covariant formulation of crosscaps in Type I
strings, see \cite{POLCHINSKI-CAI,hep-th/0009224} and for
crosscaps in asymmetric orientifold theory, see \cite{hep-th/0006033}.} . 
The crosscap state is labeled by
the spin structure $\eta$.
The above equations are valid both in the
NS-NS and the R-R sector.

In order to obtain the crosscap state as a solution to the equations
\refb{e29} it is convenient to rewrite these 
equations in terms of the oscillators. In both the NS-NS and the 
R-R sectors these equations lead to 
\begin{gather}
\label{e30b5}
(z^i_0 - e^{2\pi ikv_i}\bar{z}^i_0) \ket{C_6;p,\eta} = 0, 
\\
\label{e30b4}
(\bp^i+e^{-2\pi ikv_i}p^i) \ket{C_6;p,\eta} = 0, 
\\
\label{e30b1}
(\alpha^{\mu}_n + e^{i\pi n}\wt{\alpha}^{\mu}_{-n})
\ket{C_6;p,\eta} = 0, 
\\
\label{e30b3}
(\alpha^i_{l} +e^{-i\pi l}e^{2\pi ikv_i}
\bar{\wt{\alpha}^i}_{-l})\ket{C_6;p,\eta} = 0, 
\quad
(\bar{\alpha^i}_{l} 
+ e^{-i\pi l}e^{-2\pi ikv_i}\wt{\alpha}^i_{-l})
\ket{C_6;p,\eta} =0, 
\end{gather}
in terms of the bosonic coordinates, momenta and oscillators described in
the Appendix.
Similarly, the equations for the fermionic oscillators are,
\begin{gather}
\label{e30f1}
(\psi^{\mu}_r + i\eta e^{-i\pi r}\wt{\psi}^{\mu}_{-r})
\ket{C_6;p,\eta}=0,\\
\label{e30f2}
(\Psi^i_{r} +i\eta e^{-i\pi r}e^{2\pi ikv_i}\bar{\wt{\Psi}^i}_{-r}) 
\ket{C_6;p,\eta}=0,
\quad
(\bar{\Psi^i}_{r} +i\eta e^{-i\pi r}e^{-2\pi ikv_i}
\wt{\Psi}^i_{-r}) \ket{C_6;p,\eta}=0,
\end{gather}
where $p$ is used to designate the complex momenta 
$p^i$ in the three internal directions of the $\C^3$ and their
complex conjugates, $\bp^i$, with $i=1,2,3$ and $\eta=\pm 1$.
Equations \refb{e30b5} -- \refb{e30f2} can be solved to yield
a coherent state for the crosscap as,
\begin{equation} \label{e31}
\begin{split}
\ket{C_6;p,\eta} = 
\exp\Big( 
&-\sum_{{\mu=0,3}\atop{l\in\Z}}\frac{1}{n}(-1)^n\alpha^{\mu}_{-n}
\wt{\alpha}^{\mu}_{-n} 
-i\eta\sum_{{\mu=0,3}\atop{r> 0}} e^{-i\pi r}
\psi^{\mu}_{-r}\wt{\psi}^{\mu}_{-r} 
\\
&-\sum_{{i=1,2,3}\atop{l\in\Z_+}}
\frac{e^{i\pi l}}{l}\big(e^{-2\pi ikv_i}
\alpha^i_{-l}\wt{\alpha}^i_{-l}
+{e^{-i\pi l}}e^{2\pi ikv_i}
\bar{\alpha^i}_{-l}\bar{\wt{\alpha}^i}_{-l}\big)
\\
&-i\eta\sum_{{i=1,2,3}\atop{r>0}}
e^{-i\pi r}\big(e^{-2\pi ikv_i}\Psi^i_{-r}\wt{\Psi}^i_{-r}
+e^{2\pi ikv_i}
\bar{\Psi^i}_{-r}\bar{\wt{\Psi}^i}_{-r}
\big)\Big)
\ket{C_6; p,\eta}^{(0)}
\end{split}
\end{equation} 
where  
$\ket{C_6, p,\eta}^{(0)}$ denotes the Fock space ground 
state which is unique in the NS-NS sector 
and is independent of $\eta$. 
The ground state is, however, 
degenerate in the R-R sector and depends on $\eta$. 
Hence in considering the GSO projection and 
the orientifolding it will be convenient to
treat the NS-NS and R-R sectors separately.
\subsection*{NS-NS sector}
Let us first discuss the projections on the crosscap state 
in the NS-NS sector to obtain the invariant state.
\subsubsection{\itshape GSO Projection}
The NS-NS vacuum is chosen to be odd under the GSO projection. 
The form of the GSO projectors written in \refb{GSO} are deduced from
\begin{equation} 
\label{e34}
\begin{split}
\wi &= -(-1)^{
\sum_{r\in\Z+\half}
\left[\sum_{\mu}\psi^{\mu}_{-r}\psi^{\mu}_r
+\sum_{i=1}^3 \left(\Psi^i_{-r}\Psi^i_r + \bar{\Psi^i}_{-r}\bar{\Psi^i}_r
\right)\right]},
\\
\wir &= -(-1)^{
\sum_{r\in\Z+\half}
\left[\sum_{\mu}\wt{\psi}^{\mu}_{-r}
\wt{\psi}^{\mu}_r
+ \sum_{i=1}^3 \left(\wt{\Psi}^i_{-r}\wt{\Psi}^i_r 
+ \bar{\wt{\Psi}^i}_{-r}\bar{\wt{\Psi}^i}_r\right)\right]},
\end{split}
\end{equation} 
in terms of the oscillators in the left- and right-moving sectors. 
Their action on the ground state is given by
\begin{equation} \label{e35}
\wi\ket{p}^{(0)}_{\NSNS} = 
\wir\ket{p}^{(0)}_{\NSNS} = - \ket{p}^{(0)}_{\NSNS},
\end{equation} 
where we refrain from mentioning the spin structure explicitly, since the ground
state does not depend on it. The above equation implies
\begin{equation} \label{e36}
\wi\ket{C_6;p,\eta}_{\NSNS} = \wir\ket{C_6;p,\eta}_{\NSNS} =
-\ket{C_6;p,-\eta}_{\NSNS}.
\end{equation} 
Thus the GSO-invariant state in the NS-NS sector is,
\begin{equation} 
\label{e37}
\ket{C_6;p}_{\NSNS} = \sqthree\frac{\NN^{\NSNS}_C}{\sqrt 2}\cdot\frac{1}{\sqrt 2}
\Big[\ket{C_6;p,+}_{\NSNS} - \ket{C_6;p,-}_{\NSNS}\Big],
\end{equation} 
where $\NN^{\NSNS}_C$ is the normalization of the 
NS-NS part of the crosscap.
We have separated out a factor of ${1\over\sqrt 2}$
from the normalization factor to make sure that
it correctly reproduces the projector due to
orientifold in the open string channel. The second
factor, as usual, is for generating the correct 
GSO projector in the open channel. 
The NS-NS part of the spatially localized crosscap
is obtained by integrating this invariant
state over the internal momenta subject to the 
constraints \refb{e30b4}. Thus, the 
position eigenstate corresponding to crosscap in the untwisted NS-NS 
sector is 
\begin{equation} \label{e38}
\ket{C_6}_{\NSNS} = 
\int\prod_{i=1}^3 dp^id\bp^i\delta(p^i+e^{2\pi ikv_i}\bp^i)
\,\delta(z^i_0-e^{2\pi ikv_i}\bz^i_0)\ket{C_6;p}_{\NSNS}
\end{equation} 
\subsubsection{\itshape Orientifolding}
Considering the orientifolding on the crosscap state constructed above, 
let us first note that 
$\wil$ acts trivially on the NS-NS
ground state $\ket{p}_{\NSNS}^{(0)}$, 
which is a spacetime scalar.
The momenta and the coordinates in the internal
directions transform under orientifolding as
\begin{equation} \label{e39a}
p^i\map e^{2\pi ikv_i}\bp^i,\quad
\bp^i\map e^{-2\pi ikv_i}p^i,\quad 
z_0^i\map e^{2\pi ikv_i}\bar{z_0^i},\quad
\bar{z_0^i}\map e^{-2\pi ikv_i}z_0^i .
\end{equation} 
Since the vertex operator for the NS-NS ground state 
carrying momenta only along the internal directions 
\begin{equation} \label{e39}
e^{i\sum_{m=4}^9 P_mX^m}= e^{\frac{i}{2}\sum_{i=1}^3 (\bp^iZ^i+p^i\bar{Z^i})}
\end{equation} 
is invariant, the NS-NS ground state
$\ket{p}_{\NSNS}^{(0)}$
is invariant under the orientifolding.
Further, the oscillators transform as 
\begin{gather}
\alpha^{\mu}_n\map e^{i\pi n}\wt{\alpha}^{\mu}_n, 
\quad
\wt{\alpha}^{\mu}_n\map e^{-i\pi n}\alpha^{\mu}_n,
\quad
\psi^{\mu}_r\map e^{i\pi r}\wt{\psi}^{\mu}_r,
\quad
\wt{\psi}^{\mu}_r\map -e^{-i\pi r}\psi^{\mu}_r,
\notag\\
\alpha^i_l\map e^{i\pi l}e^{2\pi ikv_i}\bar{\wt{\alpha}^i_l},
\quad
\bar{\alpha^i_l}\map e^{i\pi l}e^{-2\pi ikv_i}\wt{\alpha}^i_l,
\quad
\wt{\alpha}^i_l\map e^{-i\pi l}e^{2\pi ikv_i}\bar{\alpha^i_l},
\quad
\bar{\wt{\alpha}^i_l}\map e^{-i\pi l}e^{-2\pi ikv_i}\alpha^i_l,
\notag
\\
\label{e41}
\Psi^i_r\map e^{i\pi r}e^{2\pi ikv_i}\bar{\wt{\Psi}^i_r},
\quad
\bar{\Psi^i_r}\map e^{i\pi r}e^{-2\pi ikv_i}\wt{\Psi}^i_r, 
\quad
\wt{\Psi}^i_r\map -e^{-i\pi r}e^{2\pi ikv_i}\bar{\Psi^i_r},
\quad
\bar{\wt{\Psi}^i_r}\map -e^{-i\pi r}e^{-2\pi ikv_i}\Psi^i_r.
\end{gather}
under orientifolding.
The exponential factor in equation \refb{e31} is invariant 
under $\g(k)$. Hence
the state $\ket{C_6;p,\eta}_{\NSNSU}$ in equation \refb{e31} 
is invariant under the 
orientifold group. The measure as well as the delta-function 
in \refb{e38} is invariant under $\g(k)$. 
Hence the crosscap state in NS-NS untwisted sector 
obtained in equation \refb{e38} is invariant under $\mathcal{G}$.
\subsection*{R-R sector}
Let us now turn to the R-R sector.
Unlike its NS-NS counterpart, the ground state in the untwisted
R-R sector is degenerate. Let us first discuss these degenerate states.
It is convenient to write the zero mode operators in the 
creation-annihilation basis of the $so(8)$ Clifford algebra.
\be\label{e47a}
\Gamma^{0,\pm} = \frac{1}{\sqrt{2}}
(\psi^0_0 \pm i\psi^3_0)\,,\qquad \Gamma^{i,\pm} = 
\frac{1}{\sqrt{2}}(\psi^{2i+2}_0 
\pm i \psi^{2i+3}_0),\quad i=1,2,3
\ee
for the left-moving states and similarly for right-moving ones, mutatis mutandis.
These  satisfy the anti-commutation relations,
\be\label{e47b}
\{\Gamma^{a,+}, \Gamma^{b,-}\} = \delta^{ab},
\quad
\{\Gamma^{a,\pm}, \Gamma^{b,\pm}\} = 0,
\ee
where $a,b=0,1,2,3$.
The untwisted R-R ground states are defined in terms of these operators as
\be\label{e47c}
(\Gamma^{0,-} +i\eta\wt{\Gamma}^{0,-})
\ket{p,\eta}^{(0)}_{\RaRa} = 0,\quad 
(\Gamma^{i,-} + i\eta\wt{\Gamma}^{i,+})\ket{p,\eta}^{(0)}_{\RaRa} 
= 0.
\ee
The ground states may also be chosen so that all the signs in $\Gamma^{\pm}$ in this
equation are reversed. However, the present choice is the one that is in harmony with 
the corresponding equations for the non-zero modes, 
\refb{e30f1}--\refb{e30f2}.
The R-R ground state is chosen to be
\be\label{e47d}
\begin{split}
\ket{C_6;p,\eta}^{(0)}_{\RaRa} =
&\exp\Big[-i\eta\big(\Gamma^{0,+}\wt{\Gamma}^{0,-}
+\sum_i\Gamma^{i,+}\wt{\Gamma}^{i,+}\big)\Big]
\ket{----}_L\ox\ket{+---}_R
\\
&-\exp\Big[-i\eta\big(\Gamma^{0,-}\wt{\Gamma}^{0,+}+\sum_i\Gamma^{i,-}
\wt{\Gamma}^{i,-}\big)\Big]
\ket{++++}_L\ox\ket{-+++}_R,
\end{split}
\ee
where $\ket{\pm,\pm,\pm,\pm}_{L,R}$ are as defined in \S~\ref{secoa} and
\be\label{e47e}
\Gamma^{a,\pm}\ket{\pm,\pm,\pm,\pm}_L =
\wt{\Gamma}^{a,\pm}\ket{\pm,\pm,\pm,\pm}_R = 0.
\ee
Let us now discuss the GSO and the orientifold projections on the
ground state and verify that it is invariant under these operations.
\subsubsection{\itshape GSO Projection}
In the R-R sector the GSO operator \refb{GSO}
acts as the chirality operator on the zero modes, namely, 
\be
\wi_{\circ}=
\Gamma_{11} = \prod_{0,3,\ldots,9}{\sqrt 2}\psi^M_0
\ee
in the left moving sector and 
\be 
\wir_{\circ}=
\wt{\Gamma}_{11} = \prod_{0,3,\ldots,9}{\sqrt 2}
\wt{\psi}^M_0
\ee
in the right-moving sector. Their respective actions on the ground states,
therefore, are given as
\be\label{e49a}
\wi_{\circ}\ket{C_6;p,\eta}_{\RaRa}^{(0)} = 
-\ket{C_6;p,-\eta}^{(0)}_{\RaRa},,\quad
\wir_{\circ}\ket{C_6;p,\eta}_{\RaRa}^{(0)} =  
\ket{C_6;p,-\eta}^{(0)}_{\RaRa}.
\ee
The form of the  GSO operator on the non-zero modes
becomes
\be\label{e51}
\begin{split}
\wi_{\bullet}&= (-1)^{\sum_{r\in\Z\setminus\{0\}}
\left[\sum_{\mu}\psi^{\mu}_{-r}\psi^{\mu}_r
+ \sum_{i=1}^3 (\Psi^i_{-r}\Psi^i_r + 
\bar{\Psi^i}_{-r}\bar{\Psi^i}_r)\right]},\\
\wir_{\bullet}&= (-1)^{\sum_{r\in\Z\setminus\{0\}}
\left[\sum_{\mu}\wt{\psi}^{\mu}_{-r}\wt{\psi}^{\mu}_r
+ \sum_{i=1}^3 (\wt{\Psi}^i_{-r}\wt{\Psi}^i_r + \bar{\wt{\Psi}^i}_{-r}
\bar{\wt{\Psi}^i}_r)\right]}.
\end{split}
\ee
They leave the  ground states unaltered.
Moreover, since the coherent state in equation \refb{e31}
contains an odd number of fermion oscillators from 
the non-zero mode sector, $\wi_{\bullet}$ and $\wir_{\bullet}$ act
only by flipping the sign of $\eta$ in the exponential.
The total GSO projection operator in the R-R sector, obtained by 
combining the two parts, namely, 
\be\label{e51a}
\wi = \wi_{\circ}\,\wi_{\bullet},
\quad 
\wir = \wir_{\circ}\,\wir_{\bullet},
\ee
act on the coherent states as,
\be\label{e51b}
\wi\ket{C_6;p,\eta}_{\RaRa} = -\ket{C_6;p,-\eta}_{\RaRa},
\quad
\wir\ket{C_6;p,\eta}_{\RaRa} = \ket{C_6;p,-\eta}_{\RaRa}.
\ee
Therefore, the GSO-invariant state in the R-R sector 
is given by, 
\be\label{e52}
\ket{C_6;p}_{\RaRa} = \sqthree\frac{\NN^{\RaRa}}{\sqrt 2}\cdot\frac{1}{\sqrt 2}
\Big[\ket{C_6;p,+}_{\RaRa} 
+ \ket{C_6;p,-}_{\RaRa}\Big],
\ee
where $\NN^{\RaRa}_C$ is the normalization of the R-R part of the crosscap.
\subsubsection{\itshape Orientifolding}
The action of the orientifold group on the R-R ground states
are given in \refb{e8}. The corresponding action
on the $\Gamma$-matrices is given by
\be\label{e52a}
\Gamma^{0,\pm}\map\wt{\Gamma}^{0,\pm},
\quad\wt{\Gamma}^{0,\pm}\map -\Gamma^{0,\pm},
\quad
\Gamma^{i,\pm}\map\wt{\Gamma}^{i,\mp},
\quad
\wt{\Gamma}^{i,\pm}\map -\Gamma^{i,\mp}.
\ee
From equation \refb{e47d} we see that the coherent state 
$\ket{C_6;p,\eta}^{(0)}_{\RaRa}$
is invariant under the orientifolding by $\GG$.
Finally, using the same measure used in equation \refb{e38} 
for constructing position the
eigenstate for NS-NS crosscap, we obtain the R-R part of  
the crosscap state
\be\label{e57}
\ket{C_6}_{\RaRa} = \int\prod_{i=1}^3 dp^id\bp^i
\delta(p^i + e^{2\pi ikv_i}\bp^i)
\,\delta(z^i_0-e^{2\pi ikv_i}\bz^i_0)\ket{C_6;p}_{\RaRa}
\ee
Finally, collecting the contributions from both the NS-NS and the R-R sectors in equation .\refb{e38} and \refb{e57} respectively, 
the crosscap state invariant under GSO projection and orientifolding is 
\be
\ket{C_6} = \frac{1}{\sqrt 2}\Big[\ket{C_6}_{\NSNS} + \ket{C_6}_{\RaRa}\Big].
\ee
This crosscap represents a canonical $O6$-plane \ie which carries negative $D6$-brane charge. The sign is fixed by choosing $\NN^{\RaRa}_C$ in \refb{e52} properly.


\section{D-brane boundary state}\label{s4}
Let us now proceed to discuss the construction of
boundary states of the D0-branes and the D-instanton in 
the present model. The boundary states for the D0-branes 
are obtained, again, by solving an appropriate set of 
boundary conditions obtained in the world-sheet theory. 
The boundary state for the D-instanton is obtained from these
by analytic continuation to an Euclidean time. 
\subsection{D0-brane}
The boundary states of D0-branes has been worked out 
in ref.\cite{hep-th/9906242} in great detail. 
We briefly review their construction here.
Let us begin with the boundary state for
the D0-branes. The boundary conditions 
satisfied by the D0-branes are
\begin{gather}
\label{e811}
\partial_{\tau}X^0(\sigma)\ket{B0} = 0,\quad
\partial_{\sigma}X^{\pm}(\sigma) \ket{B0} = 0,\quad 
\partial_{\sigma}X^m(\sigma) \ket{B0} = 0,\\
\label{e812}
(\psi^0 + i\eta\wt{\psi}^0)(\sigma) \ket{B0} = 0,\quad 
(\psi^{\pm} - i\eta\wt{\psi}^{\pm})(\sigma) \ket{B0}= 0,\quad
(\psi^m - i\eta\wt{\psi}^m)(\sigma) \ket{B0} = 0,
\end{gather}
for the space-time components of the bosonic and fermionic fields in the 
external four dimensions and 
\begin{gather}
\label{e813}
\partial_{\sigma}Z^i(\sigma) \ket{B0} = 0,\quad 
\partial_{\sigma}\bar{Z^i}(\tau=0,\sigma) \ket{B0} = 0, 
\\
\label{e814}
(\Psi^i - i\eta\wt{\Psi}^i)(\sigma) \ket{B0} = 0,\quad
(\bar{\Psi^i} - i\eta\bar{\wt{\Psi}^i}(\sigma)) \ket{B0} = 0,
\end{gather}
for the internal components, 
where $m=0,3$,~$i=1,2,3$ and we have suppressed the 
temporal coordinate $\tau$ of the world-sheet from the 
notation. All the equations are at $\tau=0$.
These equations are valid both for the NS-NS and the
R-R sector. As in the last section it is convenient to 
solve these equations in terms of the oscillators and momenta.
The coherent states in the NS-NS and the R-R
sectors are similar, except for the grading of the fermionic 
oscillators in the internal directions. 
In the NS-NS sector the index $r$ of these oscillators, $\Psi^i_r$,
are half-integral, taking values in $\Z+1/2$,
while in the R-R sector they are integral, taking values in $\Z$.
However, formulas in the untwisted and the twisted sectors are
rather different. So let us consider them in turn. 
\subsection*{Untwisted sector}
In terms of oscillators the equations for the untwisted 
NS-NS and R-R sectors become 
\begin{gather}
\label{e821}
P^0 \ket{B0} = 0,\\
\label{e822}
(\alpha^{0}_l + \wt{\alpha}^{0}_{-l}) \ket{B0} = 0,\quad
(\alpha^{\pm}_l - \wt{\alpha}^{\pm}_{-l}) \ket{B0} = 0,\quad
(\alpha^{m}_l - \wt{\alpha}^{m}_{-l}) \ket{B0} = 0,\\
\label{e823}
(\psi^0_r + i\eta\wt{\psi}^0_{-r}) \ket{B0} = 0,\quad
(\psi^{\pm}_r - i\eta\wt{\psi}^{\pm}_{-r}) \ket{B0} = 0,\quad
(\psi^m_r - i\eta\wt{\psi}^m_{-r}) \ket{B0} = 0,
\end{gather}
in the external four-dimensional part and 
\begin{gather}
\label{e824}
(\alpha^{i}_l - \wt{\alpha}^{i}_{-l})\ket{B0} = 0,\quad
(\bar{\alpha^{i}_l} - \bar{\wt{\alpha}^{i}_{-l}}) \ket{B0} = 0,
\\
\label{e825}
(\Psi^i_r - i\eta\wt{\Psi}^i_{-r}) \ket{B0} = 0,\quad
(\bar{\Psi^i_r} - i\eta\bar{\wt{\Psi}^i_{-r}}) \ket{B0} = 0.
\end{gather}
for the six-dimensional internal ones. 
From the equations \refb{e821}--\refb{e825}  we observe that 
the momenta along all the Dirichlet directions, 
including the light-cone directions, 
$P^3$, $P^{\pm}$ and the momenta of the internal 
directions $p^i$, $\bp^i$ as well as the spin structure $\eta$
are the quantum numbers labelling the boundary
states of the D0-brane and hence $\ket{B_0}$ stands for
$\ket{P^{\pm},P^3,p}_{\NSNS\atop\RaRa},\eta$ in the untwisted sector.

A coherent state for the D0-brane is obtained by solving 
\refb{e821}--\refb{e825}. The coherent states
in the untwisted NS-NS and R-R sectors are built from the respective
ground states. The ground state in the NS-NS sector
is unique, carrying momenta
$P^{\pm}$, $P^3$, $p^i$ and $\bp^i$ and is independent
of the spin structure. Thus, the contribution from the 
untwisted NS-NS sector to the coherent state is 
\be
\begin{split}
\label{e83ns}
&\ket{B0;P^{\pm},P^3,p,\eta}_{\NSNSU} = 
\exp\bigg(\sum_{l\in\Z}\frac{1}{l}
(-\alpha^0_{-l}\wt{\alpha}^0_{-l} + \alpha^3_{-l}\wt{\alpha}^3_{-l}) 
+ \sum_{{i=1,2,3}\atop{l\in\Z}}\frac{1}{l} 
(\alpha^i_{-l}\bar{\wt{\alpha}^i_{-l}} 
+ \bar{\alpha^i_{-l}}\wt{\alpha}^i_{-l})
\\&
\quad+ i\eta
\sum_{r\in\Z+\half}
(-\psi^0_{-r}\wt{\psi}^0_{-r}
+\psi^3_{-r}\wt{\psi}^3_{-r})
+ i\eta\sum_{{i=1,2,3}\atop{r\in\Z+1/2}}
(\Psi^i_{-r}\bar{\wt{\Psi}^i_{-r}}
+\bar{\Psi^i_{-r}}\wt{\Psi}^i_{-r})\bigg) 
\ket{B0;P^{\pm},P^3,p}^{(0)}_{\NSNSU},
\end{split}
\ee
where $\ket{B0;P^{\pm},P^3,p}^{(0)}_{\NSNSU}$ denotes
the ground state in the untwisted NS-NS sector.
The R-R ground states are degenerate, carrying 
momenta only along the extrenal light-cone directions and
$X^3$, contributing 
\be
\begin{split}
\label{e83rr}
&\ket{B0;P^{\pm},P^3,\eta}_{\RRU} = 
\exp\bigg(\sum_{l\in\Z}\frac{1}{l}
(-\alpha^0_{-l}\wt{\alpha}^0_{-l} + \alpha^3_{-l}\wt{\alpha}^3_{-l}) 
+ \sum_{{i=1,2,3}\atop{l\in\Z}}\frac{1}{l} 
(\alpha^i_{-l}\bar{\wt{\alpha}^i_{-l}} 
+ \bar{\alpha^i_{-l}}\wt{\alpha}^i_{-l})
\\&
\quad+ i\eta\sum_{r\in\Z} 
(-\psi^0_{-r}\wt{\psi}^0_{-r}
+\psi^3_{-r}\wt{\psi}^3_{-r})
+ i\eta\sum_{{i=1,2,3}\atop{r\in\Z}}
(\Psi^i_{-r}\bar{\wt{\Psi}^i_{-r}}
+\bar{\Psi^i_{-r}}\wt{\Psi}^i_{-r})\bigg) 
\ket{B0;P^{\pm},P^3,\eta}^{(0)}_{\RRU},
\end{split}
\ee
to the coherent state,
where $\ket{B0;P^{\pm},P^3,\eta}^{(0)}_{\RRU}$ denotes
the ground state in the untwisted R-R sector, 
given by\cite{hep-th/9906242},
\be\label{e83a}
\ket{B0;P^{\pm},P^3,p,\eta}^{(0)}_{\RRU} = 
\exp\Big[i\eta(-\Gamma^{0,+}\wt{\Gamma}^{0,+} + 
\sum_i\Gamma^{i,+}
\wt{\Gamma}^{i,-})\Big]\ket{----}_L\ox\ket{-+++}_R.
\ee
We now go on to consider the GSO projection and the orientifolding 
on the boundary states as we have done for the crosscap state in the
preceding section.
\subsubsection{\itshape GSO Projection}
Using the GSO projection operator from \refb{GSO}, we obtain the 
GSO-invariant combination of the boundary state in the
untwisted NS-NS sector as, 
\be\label{e84a}
\ket{D0;P^{\pm},P^3,p}_{\NSNSU} = \sqthree\frac{\NN^{\NSNSU}_{0}}{\sqrt 2}\cdot\frac{1}{\sqrt 2}
\Big[\ket{B0;P^{\pm},P^3,p,+}_{\NSNSU} - 
\ket{B0;P^{\pm},P^3,p,-}_{\NSNSU}\Big],
\ee
while the GSO-invariant state in the untwisted R-R sector is found to 
be 
\be\label{e84b}
\ket{D0;P^{\pm},P^3}_{\RRU} = \sqthree\frac{\NN^{\RRU}_{0}}{\sqrt 2}\cdot\frac{1}{\sqrt 2}
\Big[\ket{B0;P^{\pm},P^3,+}_{\RRU} + \ 
\ket{B0;P^{\pm},P^3,-}_{\RRU}\Big].
\ee

Finally, the position eigenstates are
obtained by integrating on the available momenta.
Assuming  these fractional branes 
to be localized at the origin of the 
internal space which is also the 
location of orbifold singularity, the 
postion eigenstates in the NS-NS and the R-R sectors are, 
respectively,
\begin{gather}
\label{e85ns}
\ket{D0}_{\NSNSU} = 
\int dP^+\ dP^-\ dP^3\prod_{i=1}^3\ dp^id\bp^i
\ket{D0;P^{\pm},P^3,p}_{\NSNSU} \\
\label{e85rr}
\ket{D0}_{\RRU} = 
\int dP^+\ dP^-\ dP^3
\ket{D0;P^{\pm},P^3}_{\RRU} 
\end{gather}
We have thus obtained the contributions to the boundary state
of the D0-brane from the untwisted NS-NS and R-R sectors.
The untwisted part of the D0-brane boundary state
in the orbifold theory is
\be\label{D0orb}
\ket{D0}^{\U}_{{\rm orb}} = \frac{1}{\sqrt 2}\Big[\ket{D0}_{\NSNSU} + \ket{D0}_{\RRU}\Big]
\ee
Having obtained the GSO-invariant untwisted state 
for the D0-brane in the orbifold theory,
let us now consider orientifolding them. 
\subsubsection{\itshape Orientifolding}
From the action given in \refb{GSO} for the states, we find 
that  the
NS-NS part remains invariant under the orientifolding. 
In the R-R sector, the 
exponential factor remains unaltered, while the 
ground state \refb{e84b} flips sign.
Hence the orientifold invariant untwisted part
is obtained by taking a linear sum
of brane and anti-brane boundary states. 
So the untwisted part of the D0-brane
boundary state in the orientifold theory
looks like
\be\label{D0orn}
\ket{D0}_{{\rm orientifold}}^{\U} = {\sqrt 2}\ \ket{D0}_{\NSNSU}
\ee
Hence it looks a like a non-BPS fractional D0-brane.

\subsection*{Twisted sectors}
Let us now consider the contributions from the twisted sectors. 
By substituting the expansions \refb{e24a} and 
\refb{e27tw} in \refb{e811} --- \refb{e814}
we obtain the boundary conditions to be satisfied by the $k$-th 
twisted sector in terms of the oscillators as,
\begin{gather}
\label{e86ns1}
P^0 \ket{B0;k} = 0\\
\label{e86ns2}
(\alpha^{\pm}_l-\wt{\alpha}^{\pm}_{-l}) \ket{B0;k} = 0,\quad 
(\alpha^{0}_l + \wt{\alpha}^{0}_{-l})\ket{B0;k} = 0,\quad
(\alpha^\mu_l - \wt{\alpha}^\mu_{-l})\ket{B0;k} = 0,\\
\label{e86ns3}
(\alpha^i_{l+kv_i} - \wt{\alpha}^{i}_{-l-kv_i})\ket{B0;k} = 0,\quad
(\bar{\alpha}^{i}_{l-kv_i}-\bar{\wt{\alpha}}^{i}_{-l+kv_i})\ket{B0;k} = 0,
\end{gather}
for the bosonic oscillators and 
\begin{gather}
\label{e86rr1}
(\psi^{\pm}_r - i\eta\wt{\psi}^{\pm}_{-r}) \ket{B0;k} = 0,\quad
(\psi^0_r + i\eta\wt{\psi}^0_{-r})\ket{B0;k} = 0,\quad
(\psi^3_r-i\eta\wt{\psi}^3_{-r})\ket{B0;k} = 0,\\
\label{e86rr2}
(\Psi^i_{r+kv_i} - i\eta\wt{\Psi}^i_{-r-kv_i})\ket{B0;k} = 0,\quad 
(\bar{\Psi}^i_{r-kv_i} - i\eta\bar{\wt{\Psi}}^i_{-r+kv_i})\ket{B0;k}=0,
\end{gather}
for the fermionic oscillators, 
where $\ket{B0;k}$ represents the boundary state in the $k$-th
twisted sector.
In the R-R sector $r$ is an integer in equations 
\refb{e86rr1} and \refb{e86rr2}, while it is half-integral in the
NS-NS sector.

A coherent
state for $\ket{B0;k}$ is constructed from the twisted
sector ground states. The twisted NS-NS and R-R 
ground states are the same as the ones used in 
constructing the crosscap state,  with  
degenerate twisted R-R ground states.
The contribution of the twisted sector $\ket{B0;k}$
to the boundary state is 
\be\label{e87ns}
\begin{split}
&\ket{B0;P^{\pm},P^3,\eta;k}_{\NSNST}= 
\exp\Bigg[\sum_{l\in\Z}\frac{1}{l}
(-\alpha^0_{-l}\wt{\alpha}^0_{-l} + 
\alpha^3_{-l}\wt{\alpha}^3_{-l})
+ i\eta\sum_{r\in\Z+1/2}
(-\psi^0_{-r}\wt{\psi}^0_{-r}
+\psi^3_{-r}\wt{\psi}^3_{-r}) 
\\
&\qquad\qquad + \sum_{{i=1,2,3}\atop {l\in\Z}}\left(\frac{1}{l-kv_i} 
\alpha^i_{-l+kv_i}\bar{\wt{\alpha}}^i_{-l+kv_i} + 
\frac{1}{l+kv_i}\bar{\alpha}^i_{-l-kv_i}
\wt{\alpha}^i_{-l-kv_i}\right)
\\
&\qquad\qquad + i\eta\sum_{{i=1,2,3}\atop{r\in\Z+1/2}}
\left(\Psi^i_{-r+kv_i}\bar{\wt{\Psi}}^i_{-r+kv_i}
+ \bar{\Psi}^i_{-r-kv_i}
\wt{\Psi}^i_{-r-kv_i}\right)\Bigg] 
\ket{B0;P^{\pm},P^3;k}^{(0)}_{\NSNST},
\end{split}
\ee
from the NS-NS sector and 
\begin{equation} \label{e87rr}
\begin{split}
&\ket{B0;P^{\pm},P^3,\eta;k}_{\RRT}= 
\exp\Bigg[\sum_{l\in\Z}\frac{1}{l}
(-\alpha^0_{-l}\wt{\alpha}^0_{-l} + 
\alpha^3_{-l}\wt{\alpha}^3_{-l})
+ i\eta\sum_{r\in\Z}
(-\psi^0_{-r}\wt{\psi}^0_{-r}
+\psi^3_{-r}\wt{\psi}^3_{-r}) \\
&\qquad\qquad + \sum_{{i=1,2,3}\atop {l\in\Z}}\left(\frac{1}{l-kv_i} 
\alpha^i_{-l+kv_i}\bar{\wt{\alpha}}^i_{-l+kv_i} + 
\frac{1}{l+kv_i}\bar{\alpha}^i_{-l-kv_i}
\wt{\alpha}^i_{-l-kv_i}\right)\\
&\qquad\qquad + i\eta\sum_{{i=1,2,3}\atop {r\in\Z}}
\left(\Psi^i_{-r+kv_i}\bar{\wt{\Psi}}^i_{-r+kv_i}
+ \bar{\Psi}^i_{-r-kv_i}
\wt{\Psi}^i_{-r-kv_i}\right)\Bigg] 
\ket{B0;P^{\pm},P^3,\eta;k}^{(0)}_{\RRT},
\end{split}
\end{equation} 
from the R-R sector. The ground states are 
\begin{gather} 
\label{e87a}
\ket{B0;P^{\pm},P^3;k=\pm 1}_{\NSNST}^{(0)} = \prod_i
\sigma^i_{\pm}\wt{\sigma}^i_{\pm}
\ket{0,\pm\frac{1}{3},\pm\frac{1}{3},\pm\frac{1}{3},}_L
\ox \ket{0,\pm\frac{1}{3},
\pm\frac{1}{3},\pm\frac{1}{3},}_R\\
\ket{B0;P^{\pm},P^3;k=\pm 1}_{\RRT}^{(0)} 
= \prod_i\sigma^i_{k,\pm}\wt{\sigma}^i_{k,\pm}
\ket{\mp\half,\pm\frac{1}{6},\pm\frac{1}{6},
\pm\frac{1}{6},}_L\ox \ket{\mp\half,\pm\frac{1}{6},
\pm\frac{1}{6},\pm\frac{1}{6},}_R
\end{gather} 
\subsubsection{\itshape GSO projection}
The GSO-invariant states are given by
\be\label{e88}
\ket{D0;P^{\pm},P^3;k}_{\NSNST\atop\RRT} = \sqthree\frac{\NN^{\T}_{\NSNST\atop\RRT}}{\sqrt 2}
\cdot\frac{1}{\sqrt 2}
\Big[\ket{B0;P^{\pm},P^3,+;k}_{\NSNST\atop\RRT} \mp
\ket{B0;P^{\pm},P^3,-;k}_{\NSNST\atop\RRT}\Big],
\ee
where the upper and lower signs are for
NS-NS and R-R sectors respectively.

The contribution to the boundary state of the 
D0-brane from the $k$-th twisted sector is given by the 
position eigenstate obtained 
by integrating over the transverse momenta as, 
\be\label{e89}
\ket{D0;k}_{\NSNST\atop\RRT} = \int dP^+\ dP^-\ dP^3\ 
\ket{D0;P^{\pm},P^3;k}_{\NSNST\atop\RRT}
\ee
So the twisted part of the D0-brane in the orbifold theory is 
\be\label{D0orbT}
\ket{D0}^{\T}_{{\rm orb}} = \frac{1}{\sqrt 2}\sum_{k=\pm 1}
\Big[\ep^k_1\ep^k_2\ket{D0;k}_{\NSNST} + \ep^k_1\ket{D0;k}_{\RRT}\Big],
\ee
where $\ep^k_i =\pm 1$ for $i=1,2$ and $k=\pm 1$, $\ep^k_1$ denotes the 
twisted R-R charges of these branes under twisted R-R field from $k$-th sector. 
Finally, $\ep^k_1\ep^k_2$ denote the phase of the untwisted NS-NS sectors
\cite{hep-th/9805019,hep-th/9910109}.
\subsubsection{\itshape Orientifolding}
The orientifolding operation keeps the part inside 
the exponential in equation .\refb{e87ns} and
\refb{e87rr} invariant. Moreover, the twisted
NS-NS ground state turns out to be invariant
in both $k=\pm 1$ sectors. However, as in the
untwisted sector, the twisted R-R sector
ground states picks up a minus sign. So the
orientifold invariant twisted part of the D0-brane 
is the sum of brane and
anti-bane system. 
\be\label{D0ornT}
\ket{D0}^{\T}_{{\rm orientifold}} = {\sqrt 2}\ 
\sum_{k=\pm 1}\ep^k_1\ep^k_2\ket{D0;k}_{\NSNST}
\ee 
Therefore, the actual boundary state of the D0-brane
surviving orientifold projection do not have any R-R part.
It is  given by the sum of \refb{D0orn} and \refb{D0ornT},
\begin{equation} \label{D0ornUT}
\begin{split}
\ket{D0}_{{\rm orientifold}} &= 
\ket{D0}^{\U}_{{\rm orientifold}} + \ket{D0}^{\T}_{{\rm orientifold}}\\
&= {\sqrt 2}\left[\ket{D0}_{\NSNSU} + 
\sum_{k=\pm 1}\ep^k_1\ep^k_2\ket{D0;k}_{\NSNST}\right]
\end{split}
\end{equation} 
This is consistent with the fact that there is no one-form
R-R field in the closed string spectrum.
\subsection{D-instanton}
In Type-IIA theory there is no BPS D-instanton state. 
We can always write down the boundary state of a
non-BPS D-instanton, which consists only of the NS-NS
part. It can be obtained by
flipping the sign in front of the timelike 
oscillators in the expression of D0 boundary state. 
The untwisted part is
\begin{equation} 
\begin{split}
\label{e9}
&\ket{D(-1),P,\eta}_{\NSNSU} =  
\exp\left[\sum_{l\in\Z\atop\mu=0,3} 
\frac{1}{l}\;\alpha^{\mu}_{-l}\wt{\alpha}^{\mu}_{-l}
+ i\eta\sum_{r\in\Z+\half\atop\mu=0,3}(-)^r \psi^{\mu}_{-r}\wt{\psi}^{\mu}_{-r} 
\right.\\ &\qquad\left.
+ \sum_{{i=1,2,3}\atop{l\in\Z_+}}
\frac{1}{l}\left(\alpha^i_{-l}\bar{\wt{\alpha}^i}_{-l} 
+\bar{\alpha^i}_{-l}\wt{\alpha}^i_{-l}\right)
+i\eta\sum_{{i=1,2,3}\atop{r\in\Z+\half}} \left(
\Psi^i_{-r}\bar{\wt{\Psi}^i}_{-r}
+ \bar{\Psi^i}_{-r}\wt{\Psi}^i_{-r}\right)
\right]\ket{B_{-1},P}^{(0)}_{\NSNS}.
\end{split}
\end{equation} 
Here, the metric along $\mu$ directions is Euclidean, 
$\delta^{\mu\nu}$, $\mu, \nu =0, 3$. Generically, in the 
orbifold theory such a D-instanton will be sourced by
twisted sector fields, so it also has a twisted NS-NS part 
\be\label{e9instw}
\begin{split}
&\ket{D(-1),P^{\pm},P^{\mu},\eta}_{\NSNST}= 
\exp\Bigg[\sum_{l\in\Z}\sum_{\mu=0,3} 
\frac{1}{l}
\alpha^{\mu}_{-l}\wt{\alpha}^{\mu}_{-l}
+ i\eta\sum_{r\in\Z+1/2}\sum_{\mu=0,3}
\psi^{\mu}_{-r}\wt{\psi}^{\mu}_{-r}) 
\\
&\qquad\qquad + \sum_{{i=1,2,3}\atop {l\in\Z}}\left(\frac{1}{l-kv_i} 
\alpha^i_{-l+kv_i}\bar{\wt{\alpha}}^i_{-l+kv_i} + 
\frac{1}{l+kv_i}\bar{\alpha}^i_{-l-kv_i}
\wt{\alpha}^i_{-l-kv_i}\right)
\\
&\qquad\qquad + i\eta\sum_{{i=1,2,3}\atop{r\in\Z+1/2}}
\left(\Psi^i_{-r+kv_i}\bar{\wt{\Psi}}^i_{-r+kv_i}
+ \bar{\Psi}^i_{-r-kv_i}
\wt{\Psi}^i_{-r-kv_i}\right)\Bigg] 
\ket{P^{\pm},P^{\mu}}^{(0)}_{\NSNST},
\end{split}
\ee
The GSO-invariant boundary state for  the  D$(-1)$-brane is 
\be\label{e9GSO}
\ket{D(-1)} = {1\over\sqrt 2}\left[\ket{D(-1)}_{\NSNSU} + 
\sum_{k=\pm 1}\ep^k_1\ep^k_2\ket{D(-1)}_{\NSNST}\right],
\ee
where $\ep^k_i = \pm 1$ for all $k-\pm 1$ and $i=1,2$, as before and
\begin{equation} \label{e9a}
\ket{D(-1)}_{\NSNSU} = \sqthree\frac{\NN_{-1}^{\U}}{\sqrt 2}\ 
\int\prod_{M=0}^9\ dP^M\ 
{1\over\sqrt 2}\Big[\ket{D(-1);P,+}_{\NSNS} - 
\ket{D(-1);P,-}_{\NSNS}\Big]
\end{equation} 
and
\begin{equation} 
\begin{split}
\quad \ket{D(-1)}_{\NSNST} = \sqthree\frac{\NN^{\T}}{\sqrt 2}
\int dP^+ dP^-dP^0dP^3\ {1\over\sqrt 2}\Big[
&\ket{D(-1), P^{\pm},P^{\mu}; +}_{\NSNST} 
\\ - &\ket{D(-1), P^{\pm},P^{\mu}; -}_{\NSNST}\Big]
\end{split}
\end{equation} 
Here $\NN_{-1}$ is the normalization factor to be 
determined in a self-consistent way later\fn{Since Type IIA
D-instanton can be obtained from a D-instanton-anti-D-instanton pair
in Type IIB and modding it out by $\wil$, 
we know that $\NN_{-1}$ is $\sqrt 2$-times
bigger than the BPS D-instanton. We shall get the same
fact by demanding that it becomes stable in
the orientifold theory.}. Obviously, 
this is not a {\em stable} D-instanton
as it contains tachyon in its spectrum.

We shall be interested in a D-instanton whose
boundary state does not have a coupling to
twisted NS-NS state\cite{Quiroz:2001xz,Braun:2002qa}. This can be obtained as
a linear combination of two D-instanton boundary states
given in equation \refb{e9GSO}, with opposite coupling
to twisted NS-NS fields for each $k$.
\be\label{Dinsnotw}
\wt{\ket{D(-1)}} = \ket{D(-1)}_{\NSNSU}
\ee

\section{One loop open string amplitudes}\label{s5}
Let us now proceed to  compute the 
annulus and M\"obius amplitudes.
The D0-brane boundary state, as given in
equation \refb{D0ornUT}, is obviously non-BPS
and hence contains tachyon in its spectrum.
Moreover, since this boundary state is a sum
of D0-$\bar{{\rm D0}}$-pair, this tachyon is a
complex field. Thequestion is whether the orientifold
can get rid of it and make it stable non-BPS D0-brane.
We don't expect that the orientifold projection can
get rid of two {\em real} tachyons, otherwise it would
be a novel feature of having a theory with both 
BPS D0-brane(bulk or non-fractional) and stable non-BPS
fractional D0-branes. Unfortunately, we could not prove
it by a direct computation in open string language. However,
boundary state analysis of this section clearly indicates
that the tachyon is {\em not} projected out in the
orientifold theory.

The absence of an R-R part in the
boundary state of the D0-branes 
simplifies the considerations as it now suffices
to consider the NS-NS amplitudes 
only. Furthermore, the crosscap state does not have
a twisted piece. Hence, for twisted
sectors the M\"obius amplitude vanishes.

For D-instanton in \refb{Dinsnotw}, however, the open string
amplitudes, annulus plus M\"obius, turns out to be free
of tachyons.

Throughout this and following sections, $t$ and $\ell$ 
will denote the tree (closed) and open (loop) channel 
Euclidean time respectively. Similarly, 
$\wt{q} = e^{-2\pi t}$ and $q = e^{-\pi\ell}$ will
denote the tree (closed) and open (loop) channel
modular parameters. The parameters $t$ and $l$ 
for various geometries are related as follows\cite{GP} 
\be\label{tl}
t =\left\{
\begin{array}{lcl}
{1}/{2\ell}\qquad\text{for Annulus},\\
{1}/{8\ell}\qquad\text{for M\"obius},\\
{1}/{4\ell}\qquad\text{for Klein bottle}.\\
\end{array}
\right.
\ee 
\subsection{D0-brane}
We first compute the annulus and the
M\"obius amplitudes associated with the 
D0-brane. As the results are different for
untwisted and twisted sectors we discuss 
them separately. The full D0-brane open string
amplitudes is a sum of annulus and M\"obius
in both untwisted and twisted sectors.
\subsection*{Untwisted sector}
\subsubsection{\itshape Annulus Amplitude}
As the D0-brane boundary states occur as
in \refb{D0orn} and \refb{e84a} we need the following 
amplitudes
\be \label{e90}
\begin{split} 
\int\limits_0^{\infty} dt\ {}_{\NSNSU}\bra{D0;+} e^{-t H_c} \ket{D0;+}_{\NSNSU} &= 
\int\limits_0^{\infty} dt\ {}_{\NSNSU}\bra{D0;-} e^{-t H_c} \ket{D0;-}_{\NSNSU} 
\\
&=
\onethird\cdot\half\cdot\frac{4\cdot(\NN_0^{\NSNSU})^2}{2} \int\limits_0^{\infty}\frac{d\ell}{2\ell^\frac{3}{2}} 
\frac{f_3(q)^8}{f_1(q)^8}
\\
\int\limits_0^{\infty} dt\ {}_{\NSNSU}\bra{D0;+} e^{-t H_c} \ket{D0;-}_{\NSNSU} &= 
\int\limits_0^{\infty} dt\ {}_{\NSNSU}\bra{D0;-} e^{-t H_c}\ket{D0;+}_{\NSNSU} 
\\
&=
\onethird\cdot\half\cdot\frac{4\cdot(\NN_0^{\NSNSU})^2}{2} \int\limits_0^{\infty}\frac{d\ell }{2\ell^\frac{3}{2}} 
\frac{f_2(q)^8}{f_1(q)^8}.
\end{split}
\ee
Using \refb{e90} we write the untwisted annulus 
amplitude of D0-brane in the orientifold
\begin{equation}  \label{e91}
\begin{split}  
\AN_{0-0}^{{\rm orientifold}} &= \int\limits_0^{\infty} dt\ {}_{{\rm orientifold}}\bra{D0}
e^{-t H_c}
\ket{D0}_{{\rm orientifold}} 
\\
&=
\onethird\cdot\half\cdot\frac{4\cdot(\NN^{\NSNSU}_0)^2}{2} 
\int\limits_0^{\infty}\frac{d\ell}{2\ell^\frac{3}{2}} 
\left(\frac{f_3(q)^8}{f_1(q)^8} - 
\frac{f_2(q)^8}{f_1(q)^8} \right).
\end{split}
\end{equation} 
Since the numerical factors appearing in the above formulas are of different
origins, we have shown them
explicitly, for example, the first factor of $\onethird$
is to reproduce the orbifold projector correctly, the second
$\half$ factor is due to the
GSO and the third one is for the orientifolding.
Comparing this expression with the orbifold case as 
given in \cite{hep-th/9906242} we can see that orientifolding
takes away the contribution from R-R sector and thus
it has a tachyonic contribution which will otherwise
be absent.
\subsubsection{\itshape M\"obius Amplitude}
The M\"obius amplitude is obtained from two  amplitudes, namely, the amplitude between
the crosscap and the D0-brane states with both having positive spin-structure, 
\begin{equation} 
\begin{split} 
\int\limits_0^{\infty} dt\ {}_{\NSNS}\bra{C_6;+}\Omega\sigmahat\wil 
e^{-t H_c} & \ket{D0;+}_{\NSNSU}\\ 
&= 
\int\limits_0^{\infty} dt\ {}_{\NSNS}\bra{C_6;-}\Omega\sigmahat\wil 
e^{-t H_c}\ket{D0;-}_{\NSNSU}
\\ 
&= \onethird\cdot\half\cdot\frac{\NN_c^{\NSNS}\cdot\NN_0^{\NSNSU}}{2}\cdot 2 \int\limits_0^{\infty}
\frac{d\ell}{2\ell^\frac{3}{2}}\;
\frac{f_3(q^2)^4}{f_1(q^2)^4},
\end{split}
\end{equation} 
and the amplitude in which the crosscap and the D0-brane have opposite spin-structures,
\begin{equation} 
\begin{split}
\int\limits_0^{\infty} dt\ {}_{\NSNS}\bra{C_6;+}\Omega\sigmahat\wil 
e^{-t H_c}&\ket{D0;-}_{\NSNSU}
\\
&= 
{}_{\NSNS}\bra{C_6;-}\Omega\sigmahat\wil 
e^{-t H_c}\ket{D0;+}_{\NSNSU} 
\\
&= \onethird\cdot\half\cdot\frac{\NN_c^{\NSNS}\cdot\NN_0^{\NSNSU}}{2}\cdot 2 \int\limits_0^{\infty}
\frac{d\ell}{2\ell^\frac{3}{2}}\;
\frac{f_3(q^2)^4}{f_1(q^2)^4}.
\end{split}
\end{equation} 
Since the D0-brane in the orientifold does not have an R-R part,
 the total untwisted D0-brane M\"obius amplitude 
\be\label{D0mobius}
\MS_{0-0} = \int\limits_0^{\infty} dt\ {}_{\NSNS}\bra{C_6}e^{-t H_c}\ket{D0}_{\NSNSU}
= \onethird\cdot\half\cdot\frac{\NN_c\cdot\NN_0}{2}\cdot 2 \int\limits_0^{\infty}
\frac{d\ell}{2\ell^\frac{3}{2}}
\Big[\frac{f_3(q^2)^4}{f_1(q^2)^4} - \frac{f_3(q^2)^4}{f_1(q^2)^4}\Big]
\ee
vanishes identically. Therefore,  
the tachyonic degree of freedom survives
and makes the D0-brane unstable in the 
untwisted sector.
\subsection*{Twisted sector}
Since there is no contribution to the M\"obius
amplitude from the twisted sectors, as the order of the orbifolding group is odd
in the present case, the
complete amplitude is given by the
annulus. Moreover, the entire contribution comes from
the twisted NS-NS sector. Using the expressions 
\begin{equation}  
\begin{split} 
\int\limits_0^{\infty} dt\ {}_{\NSNST}\bra{D0;+,k} & e^{-t H_c}\ket{D0;+,k}_{\NSNST} = 
\int\limits_0^{\infty} dt\ {}_{\NSNST}\bra{D0;-,k}e^{-t H_c}\ket{D0;-,k}_{\NSNST} 
\\
&= \onethird\cdot\half\cdot\frac{4\cdot(\NN_0^{\NSNST})^2}{2} \int\limits_0^{\infty}\frac{d\ell}{2\ell^\frac{3}{2}}\; 
\vartheta_3(0|i\ell)\;
\prod_{i=1}^3\ \sin(\pi k\nu_i)\frac{\vartheta_3(k\nu_i\ell | i\ell)}
{\vartheta_1(k\nu_i\ell | i\ell)}
\\
\int\limits_0^{\infty} dt\ {}_{\NSNST}\bra{D0;+,k} & e^{-t H_c}\ket{D0;-,k}_{\NSNST} = 
\int\limits_0^{\infty} dt\ {}_{\NSNST}\bra{D0;-,k}e^{-t H_c}\ket{D0;+,k}_{\NSNST} 
\\
&= \onethird\cdot\half\cdot\frac{4\cdot(\NN_0^{\NSNST})^2}{2}
\int\limits_0^{\infty}\frac{d\ell }{2\ell^\frac{3}{2}}\; 
\vartheta_2(0 | i\ell)\; 
\prod_{i=1}^3\ \sin(\pi k\nu_i)\frac{\vartheta_2(k\nu_i\ell | i\ell)}
{\vartheta_1(k\nu_i\ell | i\ell)}.
\end{split}
\end{equation} 
where $k=\pm1$ correspond to the 
twisted sectors, we get the annulus amplitude
\begin{equation} \label{e94}
\begin{split}
\int\limits_0^{\infty} dt\ {}_{\NSNST}\bra{D0;T;k}  e^{-t H_c}\ket{D0;T;k}_{\NSNST} 
&\\ 
=
\onethird\cdot\half\cdot\frac{4\cdot(\NN_0^{\NSNST})^2}{2}
\int\limits_0^{\infty}\frac{d\ell}{2\ell^\frac{3}{2}} 
\sum_{k=\pm1}\Bigg(
&\vartheta_3(0 | i\ell)
\prod_{i=1}^3\ \sin(\pi k\nu_i)\frac{\vartheta_3(k\nu_i\ell | i\ell)}
{\vartheta_1(k\nu_i\ell | i\ell)}
\\
-&
\vartheta_2(0 | i\ell)
\prod_{i=1}^3\ \sin(\pi k\nu_i)\frac{\vartheta_2(k\nu_i\ell | i\ell)}
{\vartheta_1(k\nu_i\ell | i\ell)}\Bigg),
\end{split}
\end{equation} 
from the twisted sector.
Expanding in $q$ we find that there are tachyonic contribution
which makes this boundary state unstable.
\subsection{D-instanton}
\subsubsection{\itshape Annulus amplitude} 
As in D0-brane case, using standard
procedure, we can work out the annulus amplitude 
for D-instanton, using \refb{e9a} and \refb{Dinsnotw} as the
boundary state.
\begin{equation} \label{e10a}
\AN_{(-1)-(-1)} = 
\int\limits_0^{\infty} dt\wt{\bra{D(-1)}}e^{-t H_c}\wt{\ket{D(-1)}} = 
\onethird\ \int\limits_0^{\infty} \frac{d\ell}{2\ell}\ 
\frac{(\NN^{\NSNSU}_{-1})^2}{2}\ 
\frac{\Big[f^8_3(q) -f_2^8(q)\Big]}{f_1^8(q)}
\end{equation} 
\subsubsection{\itshape M\"obius amplitude}
In order to calculate the M\"obius amplitude
associated with the D-instanton we need the
following expressions,
\begin{equation} \label{e24I}
\begin{split}
\int\limits_0^{\infty} dt\ {}_{\NSNS}\bra{C_6;\pm}e^{-t H_c}\ket{D(-1),\pm}_{\NSNSU} 
&= 
2\ i^{1/4}\ \int\limits_0^{\infty}\frac{d\ell}{(2\ell)^{3/2}}\ 
e^{i\pi/4}\frac{f_4^2(iq)f_3^3(q^2)}
{f_2^2(iq)f_1^3(q^2)},\\
\quad\int\limits_0^{\infty} dt\ {}_{\NSNS}\bra{C_6;\pm}
e^{-t H_c}\ket{D(-1),\mp}_{\NSNSU} 
&= 2\ i^{1/4}\int\limits_0^{\infty}\frac{d\ell}{(2\ell)^{3/2}}\ e^{-i\pi/4}
\frac{f_3^2(iq)f_3^3(q^2)}{f_2^2(iq)f_1^3(q^2)}.
\end{split}
\end{equation} 
The total M\"obius amplitude in this case is 
\begin{equation} \label{e25}
\begin{split}
\MS_{6-(-1)} + \MS_{6-(-1)}^{\ast} &= \onethird\
\int\limits_0^{\infty}\frac{d\ell}{2\ell}\ 
{1\over {2^2}}\frac{\NN_C^{\NSNS}\NN_{-1}^{\NSNSU}}{2{\sqrt 2}}\ 
\frac{2^5\cdot 2}{\sqrt 2}
i^{1/4}\Bigg[\sqtwo\Big(\frac{f_4^2(i\tq)f_3^3(\tq^2)}
{f_2^2(i\tq)f_1^3(\tq^2)} - 
\frac{f_3^2(i\tq)f_3^3(\tq^2)}
{f_2^2(i\tq)f_1^3(\tq^2)}\Big) \\
&+ {i\over\sqrt 2}\Big(\frac{f_4^2(i\tq)f_3^3(\tq^2)}
{f_2^2(i\tq)f_1^3(\tq^2)} + 
\frac{f_3^2(i\tq)f_3^3(\tq^2)}
{f_2^2(i\tq)f_1^3(\tq^2)}\Big)\Bigg]
\end{split}
\end{equation} 
\subsection{Analysis of D-instanton partition function for the tachyon}
Since the D$(-1)$-brane is a non-BPS object, 
the open string spectra on it has no GSO projection, \ie
\be\label{e28Dins}
\begin{split}
\AN_{(-1)-(-1)} + \MS_{6-(-1)} + \MS^{\ast}_{6-(-1)} = 
\onethird\cdot{1\over 2} \int\limits \frac{dl}{2l}\Bigg(\Tr_{\ns}
&\Big[(1+\Omega\sigmahat\wil )q^{H_o}
\Big] \\- &\Big[\Tr_{{\rm R}}(1+\Omega\sigmahat\wil) q^{H_o}\Big]\Bigg).
\end{split}
\ee
We can write this expression as a sum of two terms with opposite GSO 
projectors\cite{hep-th/9903123}. Thus
\be\label{e29Dins}
\AN_{(-1)-(-1)} + \MS_{6-(-1)} + \MS^{\ast}_{6-(-1)} = 
\onethird\cdot{1\over 2}\ \int\limits\frac{dl}{2l}(Z_{\ns+}(q)
+ Z_{\ns-}(q) + Z_{{\rm R}}(q)),
\ee
where
\be\label{e30Dins}
Z_{\ns\pm}(q) = \Tr_{\ns}\Big[\frac{(1+\Omega\sigmahat\wil)}{2}
(\frac{1\pm\wi}{2}) q^{H_o}\Big].
\ee 
Since the tachyon is odd under $\wi$, it lives 
in the sector with partition function $Z_{\ns-}(q)$. 
Similarly, we can know about the massless scalars, 
if we analyze $Z_{\ns+}(q)$, as the latter are $\wi$ even. 
Let us now assume that\fn{The factor $i$ in the normalization $\NN_C^{\NSNS}$ 
should not be conjugated when considering the 
conjugate crosscap state. This is a BPZ conjugation 
of CFT, not the standard quantum mechanical hermitian 
conjugation. See \cite{hep-th/9805019} for a discussion on 
this issue.} 
\be\label{Nc}
\NN^{\NSNS}_c = i x.
\ee
From the annulus and M\"obius amplitudes, 
given in equation \refb{e10a} and \refb{e25}, 
we now write down the expression for 
$Z_{\ns\pm}(q)$ and $Z_{{\rm R}}(q)$.
\begin{gather}\label{e31Dins}
Z_{\ns-}(q) = \frac{(\NN^{\NSNSU}_{-1})^2}{2} 
\frac{[f_3^8(q) + f_4^8(q)]}{2f^8_1(q)} - 
i^{1/4}\cdot\frac{x\NN^{\NSNSU}_{-1}}{2\sqrt 2}
\left[\frac{f_4^2(iq)f_3^3(q^2)}{f_2^2(iq)f_1^3(q^2)} + 
\frac{f_3^2(iq)f_3^3(q^2)}{f_2^2(iq)f_1^3(q^2)}\right],\\
Z_{\ns+}(q) = \frac{(\NN^{\NSNSU}_{-1})^2}{2}
\frac{[f_3^8(q) - f_4^8(q)]}{2f^8_1(q)} + 
i^{1/4}\cdot\frac{x\NN^{\NSNSU}_{-1}}{2\sqrt 2}
i\left[\frac{f_4^2(iq)f_3^3(q^2)}{f_2^2(iq)f_1^3(q^2)} - 
\frac{f_3^2(iq)f_3^3(q^2)}{f_2^2(iq)f_1^3(q^2)}\right],\\
Z_{{\rm R}}(q) = - \frac{(\NN^{\NSNSU}_{-1})^2}{2}\frac{f_2^8(q)}{f_1^8(q)}.
\end{gather}
The normalization of the crosscap is known 
from the compact cousin of the model under
consideration\cite{hep-th/0105208},
\be\label{Ncx}
x = 2^{5/2}.
\ee
If we choose
\be\label{insnorm}
\NN^{\NSNSU}_{-1} = 2^5,
\ee
making an expansion of these partition functions for small $q$, 
we find that the tachyon gets projected out in $Z_{\ns-}(q)$
\be\label{e32Dins}
Z_{\ns-}(q) \sim 6q + \OO(q^2).
\ee
Similarly, we find 
\be\label{e33Dins}
Z_{\ns+}(q) \sim 10 + \OO(q).
\ee 
This reflects the fact that on this D-instanton world-volume we have 
ten massless modes corresponding to the freedom of translating it along its ten
transverse dimensions.
\section{K-theory \& orientifold}\label{s6}
Having thus obtained the crosscap state and the D-brane boundary states
let us now discuss the $K$-theory
associated with the orientifold model under consideration.
The $K$-groups yield the different charges of the branes via the Chern characters.
D0-branes on orbifolds can be identified as 
objects of the derived category of an Abelian category.
The latter can be described either as the category of coherent sheaves on $\BP^2$ or as
the category of representations of the quiver associated with $\BP^2$.
These two definitions of the category are relevant in two different regimes of the
K\"ahler moduli space of the $\BP^2$. The coherent sheaves portray the D0-branes on $\BP^2$
in the large volume region, while the representations of the quiver limn
them in the orbifold limit, wherein the volume of the $\BP^2$ shrinks
\cite{hep-th/0011017,hep-th/9906242}. 
In either description, the different branes are
given by the Grothendieck group $K_0$ of the Abelian category 
\cite{sharpe,hep-th/0007175}.
For an Abelian category
$\mathcal{A}$, the equivalence classes of the objects of $\mathcal{A}$
modulo the relation $[X'] = [X]+[X'']$, when the objects $X,X',X''$ form a short exact
sequence, $\xymatrix{0\ar[r] & X\ar[r] & X'\ar[r] & X''\ar[r] &0}$, form an Abelian
group, called the Grothendieck group, denoted $K_0$, of $\mathcal{A}$. Here $[X]$ 
denotes the class of an object in $\mathcal{A}$. The equivalence relation embodies the
identification of anti-branes as objects shifted by a unit grade in a complex relative
to the objects corresponding to the branes. 

After orientifolding, however, the Grothendieck group falls inadequate to describe the
charges of the objects of the Abelian category, $\mathcal{A}$. The charges now are given
by the  Whitehead group of $\mathcal{A}$, denoted $K_1$. The classes of objects in the
$K_1$ group carry the information of certain automorphisms of the objects in addition to
the objects themselves. Before presenting the definition of the $K_1$ group, let us note
the following from our earlier discussions. We found that the boundary states of 
the D0-branes are invariant under the orientifolding operation. Further, acting twice,
the orientifolding yields the original state back. Thus, given that each boundary state
corresponds to an object in the Abelian category $\mathcal{A}$, it is natural to assume
the existence of an automorphism, $\mathfrak{f}$ of the objects of $\mathcal{A}$, such that
$\mathfrak{f}^2=1$. Since an Abelian category  is
idempotent complete and hence so is its bounded derived category \cite{balmer}, this
is a consistent assumption.
In order to interpret this automorphism physically, let us note that 
the CFT analysis points at the stable object on the orientifold backgrounds
being a D-instanton which are to be obtained through tachyon condensation
of the D0-brane. The latter was found to have tachyonic modes.  
This bears a close resemblance
to the orientifold of the Type-IIB theory, where the orientifold action is a lone
$\wil$ \cite{hep-th/0007175,hep-th/9812135, hep-th/9910109}. In this case
a D8-brane appears through tachyon condensation
of a D9-brane. From this point of view the D8-brane
has been interpreted as a pair consisting of
a vector bundle $V$ and an automorphism 
$\alpha : V \rt V$, where $V$ and $\alpha$ 
is taken to corresponds to the D9-brane and the tachyon, respectively.
In the spirit of this analogy, in the Type-IIA orientifold 
we can identify the tachyons associated with
a D0-brane described by an object $X$ as 
an automorphism $\mathfrak{f} : X \rt X$ and
D-instantons as a pair $(X,\mathfrak{f})$.
Thus, the classes of D-branes relevant for our discussion will carry an extra label
designating the automorphism $\mathfrak{f}$. This leads us to consider the Whitehead group
$K_1$ of the Abelian category  \cite{lam,hep-th/9812135}.
The Whitehead group $K_1$ of the category $\mathcal{A}$ is 
defined to be an Abelian group generated by 
equivalence classes $[X,\alpha]$, corresponding to the 
objects $X$ of $\mathcal{A}$, $\alpha$ being an automorphism of $X$. The equivalence
relations are,
\begin{dinglist}{109}
\item \emph{Additive:} For any commutative diagram
\begin{equation} 
\begin{split}
\xymatrix{
0\ar[r] & X'\ar[r]\ar[d]_{\alpha'}& X\ar[r]\ar[d]_{\alpha}
& X''\ar[r]\ar[d]_{\alpha''} & 0 \\
0\ar[r] & X'\ar[r]& X\ar[r]& X''\ar[r] & 0 
}
\end{split}
\end{equation} 
in $\mathcal{A}$, where the horizontal sequences are 
exact and the vertical morphisms  are automorphisms,
we impose an equivalence relation, 
\be
[X,\alpha] = [X',\alpha'] + [X'' , \alpha'']
\ee 
\item \emph{Multiplicative:}
If $\alpha: X\rt X$ and $\beta: X \rt X$
are automorphisms in of objects in $\mathcal{A}$, then 
we impose 
\begin{equation} 
\label{eq:mult}
[X, \alpha\beta] = [X,\alpha] + [X,\beta].
\end{equation} 
\end{dinglist}
Now, for the case at hand, the orientifolding operation lifts up to an automorphism 
$\mathfrak{f}$ on $\mathcal{A}$, as alluded to above. Then, from equation \refb{eq:mult} we
obtain, $ 2 [X,\mathfrak{f} ] = [X,1],$ which in turn yields 
\[[X,\mathfrak{f} ] = 0 \mod 2,\] 
since, again by equation \refb{eq:mult}, we have $[X,1]=0$, which can be seen
by setting $\alpha=\beta=1$.
We conclude, therefore, that the Whitehead group $K_1$ of the category of objects in the
orientifold background is isomorphic to $\Z_2$. The $\Z_2$ charge corresponding to
$K_1\sim\Z_2$ is topological and leads to the existence of a single $\Z_2$-charged
D-instanton, which we found through the analysis of the boundary states above.
\section{Discussions and conclusion}\label{s7}
To conclude, we have discussed 
the boundary states in the spectrum of the Type-IIA
orientifold $\C^3/\Z_3\cdot\Omega\cdot\sigmahat\cdot\wil$. We find that there is no
vector multiplet in the closed string
massless spectrum. Further, the D0-branes
are inflicted with tachyon and become
unstable.
However, we found the existence of a D-instanton which is stable
as the associated tachyons are projected
out due to orientifolding.

In spite of the presence of a massless modulus in the closed string spectrum of the
twisted NS-NS sector, in this article we refrain from presenting a geometric
interpretation of the objects in the aforementioned Abelian category, as    
the contribution of instantons may develop a superpotential for this modulus
\cite{vafaetal} 
rendering the movement over the K\"ahler moduli space obstructed. 

We interpret the D-instanton alluded to above as being obtained
through tachyon condensation of the unstable D0-branes.
We use the analogy with the mechanism
by which D8-branes come into being through tachyon
condensation of D9-branes.
The associated $K$-group is given by
Whitehead group of the 
category of D0-branes 
generated by objects along with an
automorphism of the object
associated with the tachyon.
Contrary to the cases studied earlier in the literature, we assume a lift of the
orientifolding on the objects of the Abelian category of D-branes, rather than looking
for a geometric realization the target space. 
One advantage of this description
of the $K$-group is that they are readily 
generalizable to more exotic orbifolds and, perhaps, to the Calabi-Yau spaces. 

However, let us note that the other way to look at the
$K$-group is \cite{hep-th/0007175}
by considering the fact that the operation
$\Omega\cdot\sigmahat\cdot\wil$
corresponds to interchanging
$(E,F)$ with $({\bar F}, {\bar E})$
where $E$ and $F$ are vector bundles
and ${\bar E}$ represents the complex 
conjugate of $E$. 
The $K$-group is given by $KR^1_{\pm}(X)$ \cite{hep-th/0007175}. 
This has not been discussed much in the literature.
It will be interesting to compute this and compare with the results here.
\bigskip

\bec
{\large\bf Acknowledgments}
\eec
We would like to thank E.~Keski-Vakkuri, 
E.~Kiritsis, A.~Sen,
for helpful discussion. JM would like to thank
Ralph Blumenhagen and Stefan F\"orste for several 
useful email correspondences. JM and SM thank 
the Department of Theoretical Physics,
Indian Association for the Cultivation of Science, 
India, where part of this work was done.
We thank the referee of this paper for pointing out inaccuracies in section \ref{s3}  
in the earlier version.
\appendix{Notations \& conventions}\label{appendix}
Here we explain the notations and conventions used in this article and present an
inventory of the formulas used.
We use light-cone gauge throughout our 
discussion and double wick-rotated the coordinates, as is customary. 
Thus, $X^0\rt iX^0$ and
$X^1\rt iX^1$. Similarly $\psi^0\rt i\psi^0$ 
and $\psi^1\rt i\psi^1$. 
The light-cone coordinates are taken to be 
$X^{\pm}= X^1\pm X^2$ and $\psi^{\pm} =
\psi^1 \pm \psi^2$. 
We impose Dirichlet boundary conditions 
along the light-cone directions.
The \emph{external} directions will be $X^0$ and $X^3$; 
both of which are taken to be
euclidean, \ie the external metric is 
$\delta^{\mu\nu}$. The coordinates 
$X^4,\ldots,X^9$ will be used to denote the \emph{internal}
$\C^3$ directions. We use indices
$M,N=0,\ldots,9$ for the ten-dimensional spacetime directions
and $\wh{M}=0,3,\ldots,9$ for indices excluding 
the light-cone directions,
$\mu,\nu=0,3$ for external indices, 
$m,n=4,\ldots,9$ for internal real 
indices and $i,\bi=1,2,3$ for complexified 
internal indices. The worldsheet
coordinates will be denoted by $\sigma$ and $\tau$ 
and we use a Euclidean metric
on the worldsheet.
\subsection{Mode Expansion} 
\subsubsection{Bosonic Oscillators} 
\subsubsection*{Light cone directions} 
\begin{equation} 
\label{e21a}
\begin{split}
X^-(\sigma,\tau) &= x^-_0 + \frac{1}{2\pi}{P^-\tau} + 
\frac{i}{2}\sum_{l\in \Z}\frac{1}{l}\alpha^-_l e^{-il(\tau+\sigma)}
+ \frac{i}{2}\sum_{l\in\Z}\frac{1}{l}\wt{\alpha}^-_le^{-il(\tau-\sigma)},\\
X^{+}(\sigma,\tau) &= x^+_0 + P^+ \tau,
\end{split}
\end{equation} 
the latter implying 
\begin{equation} 
\alpha^+_0 = \wt{\alpha}^+_0 = P^+\delta_{n0}\,,
\qquad \alpha^-_0 + \wt{\alpha}^-_0 = P^-
\end{equation} 

\subsubsection*{Untwisted sector} 
The mode expansion of the Bosonic fields are
\begin{equation} 
\label{e22}
\begin{split}
X^M(\sigma,\tau) &= x_0^M + 2P^M\tau + 
\frac{i}{2}\sum_{l\in \Z}\frac{1}{l}\alpha^M_l e^{-il(\tau+\sigma)}
+ \frac{i}{2}
\sum_{l\in\Z}\frac{1}{l}\wt{\alpha}^M_le^{-il(\tau-\sigma)},\\ 
\end{split}
\end{equation} 
For $m=4,\ldots,9$, let
\begin{equation} \label{e23}
\frac{1}{\sqrt 2}(X^{2i+2} + iX^{2i+3}) = Z^i\quad
\frac{1}{\sqrt{2}}(P^{2i+2} + iP^{2i+3}) = p^i\qquad i=1,2,3
\end{equation} 
and similarly for right-movers, mutatis mutandis.
Thus the mode expansions are given as
\begin{equation} 
\label{e24}
\begin{split}
Z^i(\sigma,\tau) &= z^i_0 + 2p^i\tau +
\frac{i}{2}\sum_{l\in\Z}\frac{1}{l}\alpha^i_l e^{-il(\tau+\sigma)}
+ \frac{i}{2}\sum_{l\in\Z}\frac{1}{l}\wt{\alpha}^i_l e^{-il(\tau-\sigma)},\\
\bar{Z}^i(\sigma,\tau) &= \bz^i_0 + \bp^i\tau +
\frac{i}{2}\sum_{l\in\Z}\frac{1}{l}\bar{\alpha}^i_l e^{-il(\tau+\sigma)}
+ \frac{i}{2}\sum_{l\in\Z}\frac{1}{l}\bar{\wt{\alpha}}^i_l 
e^{-il(\tau-\sigma)}
\end{split}
\end{equation} 
\subsubsection*{Twisted sectors} 
The mode expansions in the twisted sectors, cooresponding to $k=1,2$ are 
\begin{equation} 
\label{e24a}
\begin{split}
X^{\mu}(\sigma,\tau) &= \frac{i}{2}\sum_{l\in\Z}
\frac{1}{l}\alpha^{\mu}_{l} e^{-il(\tau+\sigma)}
+ \frac{i}{2}\sum_{l\in\Z}\frac{1}{l}\wt{\alpha}^{\mu}_{l} 
e^{-il(\tau-\sigma)},\\
Z^i(\sigma,\tau) &=\frac{i}{2}\sum_{l\in\Z}\ 
\frac{1}{l+kv_i}\alpha^i_{l+kv_i} e^{-i(l+kv_i)(\tau+\sigma)}
+\frac{i}{2}\sum_{l\in\Z}\frac{1}{l-kv_i}\wt{\alpha}^i_{l-kv_i} 
e^{-i(l-kv_i)(\tau-\sigma)},\\
\bar{Z}^i(\sigma,\tau) &=\frac{i}{2}\sum_{l\in\Z}
\frac{1}{l-kv_i}\bar{\alpha}^i_{l-kv_i} e^{-i(l-kv_i)(\tau+\sigma)}
+\frac{i}{2}\sum_{l\in\Z}\frac{1}{l+kv_i}\bar{\wt{\alpha}}^i_{l+kv_i} 
e^{-i(l+kv_i)(\tau-\sigma)}.
\end{split}
\end{equation} 
\subsubsection{Fermionic Oscillators} 
\subsubsection*{Light-cone directions}
\begin{alignat}{2}
\label{e24b}
\psi^+(\sigma,\tau) = \wt{\psi}^+(\sigma,\tau) = 0
&\quad\imply\quad\psi^+_r =
\wt{\psi}^+_r = 0,\\
\psi^-(\sigma,\tau) = \frac{2}{P^+}\ 
\sum_{\wh{M}=0,3,\ldots,9}\psi^{\wh{M}}\partial_+ X^{\wh{M}}
&\quad\imply\quad\psi^-_r = \frac{1}{P^+}
\sum_{\wh{M}=0,3,\ldots,9}\sum_s \alpha^{\wh{M}}_{r-s}\psi^{\wh{M}}_s,
\\
\wt{\psi}^-(\sigma,\tau) = \frac{2}{P^+} 
\sum_{\wh{M}=0,3,\ldots,9}\wt{\psi}^{\wh{M}}\partial_- X^{\wh{M}}
&\quad\imply\quad\wt{\psi}^-_r = \frac{1}{P^+}
\sum_{\wh{M}=0,3,\ldots,9}\sum_s \wt{\alpha}^{\wh{M}}_{r-s}
\wt{\psi}^{\wh{M}}_s,
\end{alignat}
where $r,s\in\Z+\half(\Z)$ for NS(R)-sector and $\sigma^{\pm} = \tau \pm
\sigma$, $\partial_{\pm}  = \half(\partial_{\tau} \pm \partial_{\sigma})$.
\subsubsection*{Untwisted sector} 
The mode expansions of the left- and right-moving fermions are
\begin{equation} \label{e25U}
\psi^M(\sigma,\tau) = \sum_r \psi^M_r e^{-ir(\tau +\sigma)}
\quad\text{and}\quad 
\wt{\psi}^M(\sigma,\tau) = \sum_r \wt{\psi}^M_r
e^{-ir(\tau -\sigma)},
\end{equation} 
respectively, where $r\in\Z+\half$($\Z$) for NS(R)-sectors and $m=4,\ldots,9$. 
We define complex fermionic fields, 
\begin{equation} 
\Psi^i = \frac{1}{\sqrt 2}(\psi^{2i+2} +i\psi^{2i+3}),\qquad
\bar{\Psi^i} = \frac{1}{\sqrt 2}(\psi^{2i+2} -i\psi^{2i+3}),
\end{equation} 
where $i=1,2,3$ 
and similarly $\wt{\Psi}^i$ and $\bar{\wt{\Psi}^i}$ for right-movers. The
mode expansion for the complexified fermions can be derived from the earlier expressions
as 
\begin{equation} 
\begin{split}
\label{e27}
\Psi^i(\sigma,\tau) = \sum_r \Psi^i_r e^{-ir(\tau +\sigma)}, &\quad
\bar{\Psi^i}(\sigma,\tau) = \sum_r \bar{\Psi^i_r} e^{-ir(\tau+\sigma)}\\
\wt{\Psi}^i(\sigma,\tau) = \sum_r \wt{\Psi}^i_r e^{-ir(\tau-\sigma)},&\quad
\bar{\wt{\Psi}^i}(\sigma,\tau) = \sum_r \bar{\wt{\Psi}^i_r}
e^{-ir(\tau-\sigma)}
\end{split}
\end{equation} 
\subsubsection*{Twisted sectors}
The mode expansions of the fermions in the $k=1,2$ twisted sectors are
\begin{gather}
\psi^{\mu}(\sigma,\tau) = \sum_r \psi^{\mu}_{r} 
e^{-ir(\tau+\sigma)},\quad
\wt{\psi}^{\mu}_r(\sigma,\tau) = \sum_r \wt{\psi}^{\mu}_{r}
e^{-ir(\tau -\sigma)},\notag\\
\label{e27tw}
\Psi^i(\sigma,\tau) = \sum_r \Psi^i_{r+kv_i} 
e^{-i(r+kv_i)(\tau+\sigma)},\quad
\bar{\Psi^i}(\sigma,\tau) = \sum_r \bar{\Psi^i}_{r-kv_i} 
e^{-i(r-kv_i)(\tau +\sigma)},\\
\wt{\Psi}^i(\sigma,\tau) = \sum_r \wt{\Psi}^i_{r-kv_i} 
e^{-i(r-kv_i)(\tau-\sigma)},\quad
\bar{\wt{\Psi}^i}(\sigma,\tau) = \sum_r \bar{\wt{\Psi}^i}_{r+kv_i} 
e^{-i(r+kv_i)(\tau-\sigma)}. \notag
\end{gather}
\subsubsection*{Oscillator algebra}
We use the following commutators and anti-commutators for the oscillators
\be\label{e27b}
[\alpha^{\mu}_l,\alpha^{\nu}_{l'}] = 
[\wt{\alpha}^{\mu}_l,
\wt{\alpha}^{\nu}_{l'}] = \delta_{l+l',0}\,\delta^{\mu\nu},\qquad
\{\psi^{\mu}_r,\psi^{\nu}_s\} = \{\wt{\psi}^{\mu}_r,\wt{\psi}^{\nu}_s\}
= \delta_{r+s,0}\,\delta^{\mu\nu},
\ee
for the external directions and 
\begin{alignat}{2}
[\alpha^i_{l+kv_i},\bar{\alpha}^j_{l'-kv_j}]
= (l+kv_i)\delta_{l+l',0}\,\delta^{ij},
&\quad\{\Psi^i_{r+kv_i},\bar{\Psi}^j_{s-kv_j}\} = 
\delta_{r+s,0}\,\delta^{ij}, \\
[\wt{\alpha}^i_{l-kv_i},\bar{\wt{\alpha}}^j_{l'+kv_j}]
= (l-kv_i)\delta_{l+l',0}\,\delta^{ij},
&\quad\{\wt{\Psi}^i_{r-kv_i}, \bar{\wt{\Psi}}^j_{r+kv_j}\} = 
\delta_{r+s,0}\,\delta^{ij},
\end{alignat}
for the internal directions.
\subsection{Closed string Hamiltonian}
The closed string Hamiltonian in the untwisted sector is
\begin{equation} 
\label{e27c}
\begin{split}
H^u_c &= \pi(P^2 + p^2) + 2\pi\Big[
\sum_{{\mu=0,3}\atop {n\in\Z_+}}(\alpha^{\mu}_{-n}\alpha^{\mu}_n 
+\wt{\alpha}^{\mu}_{-n}\wt{\alpha}^{\mu}_n) 
+ \sum_{{\mu=0,3}\atop{r>0}} r(\psi^{\mu}_{-r}\psi^{\mu}_r 
+\wt{\psi}^{\mu}_{-r}\wt{\psi}^{\mu}_r)\\ 
&\quad +\sum_{{i=1,\ldots,3}\atop {n\in\Z_+}}(\alpha^i_{-n}\bar{\alpha}^i_n 
+\wt{\alpha}^i_{-n}\bar{\wt{\alpha}}^i_n )
+ \sum_{{i=1,\ldots,3}\atop{r>0}} r(\Psi^i_{-r}\bar{\Psi}^i_r  
+ \wt{\Psi}^i_{-r}\bar{\wt{\Psi}}^i_r)\Big] 
+ 2\pi a_u\,,
\end{split}
\end{equation} 
where $a_u = -1$ in the NS-sector and $a_u=0$ in the R-R-sector.
The fermionic mode-index $r\in\Z+\half(\Z)$ for NS(R)-sector 
and $\vec{P}$ and $\vec{p}$ are denote the external and internal momenta, respectively.
In the twisted sector, on the other hand, the closed string Hamiltonian assumes the form
\begin{equation} 
\label{e27d}
\begin{split}
H^T_c &= \pi P^2  + 2\pi\Big[(\sum_{{\mu=0,3}\atop{n\in\Z_+}}(\alpha^{\mu}_{-n}\alpha^{\mu}_n 
+\wt{\alpha}^{\mu}_{-n}\wt{\alpha}^{\mu}_n) 
+ \sum_{{\mu=0,3}\atop{r>0}} r(\psi^{\mu}_{-r}\psi^{\mu}_r  
+\wt{\psi}^{\mu}_{-r}\wt{\psi}^{\mu}_r)\\ 
&\quad+\sum_{{i=1,\ldots,3}\atop {n\in\Z}}
\norder(\alpha^i_{n+kv_i}\smash{\bar{\alpha}^i_{-n-kv_i}}
+\wt{\alpha}^i_{n-kv_i}\bar{\wt{\alpha}}^i_{-n+kv_i})\norder\\
&\quad+ \sum_{{i=1,\ldots,3}\atop {r>0}}(r-kv_i)\norder(\Psi^i_{r+kv_i}\bar{\Psi}^i_{-r-kv_i}  
+ \wt{\Psi}^i_{r-kv_i}\bar{\wt{\Psi}}^i_{-r+kv_i} )\norder\Big] + 2\pi a_T,
\end{split}
\end{equation} 
where $a_T$ is a constant arising from the normal ordering.
\subsection{The $\vartheta$-functions and the $f$-functions} 
Let us list the $\vartheta$-functions with characteristics for 
convenience\cite{Polchinski:1998rq},
\be\label{e93char}
\vartheta\left[\alpha\atop\beta\right]\left(\nu\Big|\tau\right) = 
e^{2\pi i\alpha\beta} \hq^{\alpha^2/2 - 1/24}\ \eta(\tau)\
\prod_{n=1}^{\infty}\left(1 + \hq^{n - 1/2 + \alpha}e^{2\pi i(\beta + \nu)}\right)
\left(1 + \hq^{n - 1/2 - \alpha} e^{-2\pi i(\beta + \nu)}\right),
\ee
where $\hq= e^{2\pi i\tau}$ and one needs to 
choose $\alpha,\beta\in (-1/2,1/2]$. Here $\eta(\tau)$ 
is the Dedekind $\eta$-function. 
\be\label{eta}
\eta(\tau) = \hq^{1/24}\prod_{n=1}^{\infty} (1 - \hq^n).
\ee
The Jacobi $\vartheta$-functions are given by
the following $\vartheta$-functions with 
characteristics
\be
\begin{split}
\vartheta\left[\half\atop\half\right](\nu|\tau) &= 
\vartheta_1(\nu|\tau),\qquad 
\vartheta\left[\half\atop 0\right](\nu|\tau) = \vartheta_2(\nu|\tau),\\
\vartheta\left[0\atop 0\right](\nu|\tau) &= \vartheta_3(\nu|\tau),\qquad 
\vartheta\left[0\atop\half\right](\nu|\tau) = \vartheta_4(\nu|\tau).
\end{split}
\ee
We also use the following $f$-functions, 
related to the $\vartheta$-functions\cite{POLCHINSKI-CAI,GP} 
\begin{alignat}{2}
&f_1(\hq) = \hq^{1/12}\prod_{n=1}^\infty (1-\hq^{2n}),&\quad
f_2(\hq) = {\sqrt 2} \hq^{1/12}\prod_{n=1}^\infty (1 + \hq^{2n}),\\
&f_3(\hq) = \hq^{-1/24}\prod_{n=1}^\infty (1 + \hq^{2n-1}),&\quad 
f_4(\hq) = \hq^{-1/24}\prod_{n=1}^\infty (1 - \hq^{2n-1}).
\end{alignat}
\subsection{Modular $S$ Transformation}
The behavior of the 
Dedekind's $\eta$-function and the Jacobi $\vartheta$-functions under the modular $S$
transformation is given by,
\begin{gather} 
\label{jacobimod}
\eta\left(-\frac{1}{\tau}\right) = (-i\tau)^{-1/2}\eta(\tau),\\
\vartheta_1\left(\frac{\nu}{\tau}|-\frac{1}{\tau}\right) 
= -(-i\tau)^{1/2}e^{\pi i\nu^2/\tau} \vartheta_1 (\nu|\tau),\\
\vartheta_2\left(\frac{\nu}{\tau}|-\frac{1}{\tau}\right) 
= (-i\tau)^{1/2}e^{\pi i \nu^2/\tau} \vartheta_4 (\nu|\tau),\\
\vartheta_3\left(\frac{\nu}{\tau}|-\frac{1}{\tau}\right)
= (-i\tau)^{1/2}e^{\pi i \nu^2/\tau} \vartheta_3 (\nu|\tau),\\
\vartheta_4\left(\frac{\nu}{\tau}|-\frac{1}{\tau}\right)
= (-i\tau)^{1/2}e^{\pi i\nu^2/\tau}\vartheta_2 (\nu|\tau).
\end{gather} 
The modular $S$ transformation for the $f$-functions 
for a real (used for annulus) and an imaginary (used for M\"obius) arguments are,
\subsubsection{Real Arguments}
\be\label{freal}
\begin{split}
f_1(e^{-\pi s}) &= \frac{1}{\sqrt s}f_1(e^{-\pi/s}),\qquad
f_2(e^{-\pi s}) = f_4(e^{-\pi/s}),\\
f_3(e^{-\pi s}) &= f_3(e^{-\pi/s}),\qquad
f_4(e^{-\pi s}) = f_2(e^{-\pi/s}).
\end{split}
\ee
\subsubsection{Imaginary Arguments}
\be\label{fimg}
\begin{split}
f_1(ie^{-\pi s}) &= \frac{1}{{\sqrt 2s}}f_1(ie^{-\pi/4s}),\qquad
f_2(ie^{-\pi s}) = f_2(ie^{-\pi/4s}),\\
f_3(ie^{-\pi s}) &= e^{i\pi/8} f_4(ie^{-\pi/4s}),\qquad
f_4(ie^{-\pi s}) = e^{-i\pi/8} f_3(ie^{-\pi/4s}).
\end{split}
\ee

\end{document}